\documentclass[10pt,journal]{IEEEtran}

\IEEEoverridecommandlockouts
\ifCLASSOPTIONcompsoc
    \usepackage[nocompress]{cite}
\else
    \usepackage{cite}
\fi
\ifCLASSOPTIONcaptionsoff
    \usepackage[nomarkers]{endfloat}
\let\MYoriglatexcaption\caption
\renewcommand{\caption}[2][\relax]{\MYoriglatexcaption[#2]{#2}}
\fi
\usepackage[T1]{fontenc}
\usepackage{url}
\usepackage{subfigure}
\usepackage{dsfont}
\usepackage{bm}
\usepackage[pdftex]{graphicx}
\usepackage{array,amsmath,amssymb,amsfonts}
\usepackage{amsthm}
\usepackage{subeqnarray}
\usepackage{cases}
\usepackage{booktabs}
\usepackage{color}
\usepackage[ruled,lined,linesnumbered,commentsnumbered,longend]{algorithm2e}
\makeatletter
\newcommand{\removelatexerror}{\let\@latex@error\@gobble}
\makeatother

\SetCommentSty{mycommfont}
\renewcommand{\vec}[1]{\boldsymbol{#1}}

\DeclareMathOperator*{\argmax}{argmax}

\newtheorem{theorem}{\textbf{Theorem}}

\newtheorem{proposition}{\textbf{Proposition}}
\newtheorem{definition}{\textbf{Definition}}
\def\BibTeX{{\rm B\kern-.05em{\sc i\kern-.025em b}\kern-.08em
    T\kern-.1667em\lower.7ex\hbox{E}\kern-.125emX}}
\vspace{-0.3cm}
\IEEEoverridecommandlockouts

\usepackage{siunitx} 

\begin{document}

\title{Theoretically Guaranteed Online Workload Dispatching for Deadline-Aware Multi-Server Jobs}

\author{
        Hailiang~Zhao,
        Shuiguang~Deng,~\IEEEmembership{Senior~Member,~IEEE,} 
        Jianwei~Yin,
        Schahram~Dustdar,~\IEEEmembership{Fellow,~IEEE},
        and~Albert~Y.~Zomaya,~\IEEEmembership{Fellow,~IEEE}%
\IEEEcompsocitemizethanks{
  }%
}

\IEEEtitleabstractindextext{%
\begin{abstract}
    Multi-server jobs are imperative in modern computing clusters. A multi-server job has multiple task components and each of 
    the task components is responsible for processing a specific size of workloads. Efficient online workload dispatching is 
    crucial but challenging to co-located heterogeneous multi-server jobs. The dispatching policy should decide $(i)$ where to 
    launch each task component instance of the arrived jobs and $(ii)$ the size of workloads that each task component processes. 
    Existing policies are explicit and effective when facing service locality and resource contention in both offline and online 
    settings. However, when adding the deadline-aware constraint, the theoretical superiority of these policies could not be 
    guaranteed. To fill the theoretical gap, in this paper, we design an $\alpha$-competitive online workload dispatching policy 
    for deadline-aware multi-server jobs based on the spatio-temporal resource mesh model. We formulate the problem as a social 
    welfare maximization program and solve it online with several well designed pseudo functions. The social welfare is formulated 
    as the sum of the utilities of jobs and the utility of the computing cluster. The proposed policy is rigorously proved to be 
    $\alpha$-competitive for some $\alpha \geq 2$. We also validate the theoretical superiority of it with simulations and the 
    results show that it distinctly outperforms two handcrafted baseline policies on the social welfare.
\end{abstract}

}

\maketitle

\IEEEdisplaynontitleabstractindextext

\ifCLASSOPTIONpeerreview
\begin{center} \bfseries EDICS Category: 3-BBND \end{center}
\fi
%
\IEEEpeerreviewmaketitle

\section{Introduction}\label{s1}

\IEEEPARstart{T}oday's computing clusters are full of multi-server jobs. A multi-server job is composed of multiple associated 
task components, and each task component may take different size of input workloads and require various kinds and quantities of 
computation resources such as CPUs, GPUs, FPGAs, etc. A typical multi-server job is the distributed training of deep neural networks. 
Based on the Ring All-Reduce communication pattern\footnote{Another widely used distributed training architecture is the Parameter 
Server (PS)-Worker architecture, which will not be the focus of the proposed algorithm in this work.}, which is well supported by 
the NVIDIA Collective Communication Library (NCCL) \cite{nccl}, each task component is responsible for processing the input data 
workloads (update its local gradients with the input mini-batch data samples and take the reduction operation) and commissioning 
the NCCL library to send the reduced gradient chunks to the other task components.

Efficient workload dispatching is crucial but challenging to co-located heterogeneous multi-server jobs. There are three key problems 
need to be carefully addressed. Firstly, for each multi-server job, how many task component instances should be launched? For distributed 
model training, the more workers we start, the faster the training speed\footnote{Although more workers lead to more communication overhead, 
the overall process runs faster \cite{DL2}.}. Secondly, which node to choose to launch each task component instance? Thirdly, how many 
workloads should each task component processes? The major challenges are discussed as follows.
\begin{itemize}
    \item \textit{Service Locality.} Service locality is common in modern computing clusters. With this constraint, the task components 
    of a multi-server job can only be processed by a subset of nodes where the resource requirements, affinity \& anti-affinity 
    \cite{carrion2022kubernetes}, and other obligatory constraints are satisfied. For a resource-constrained cluster, service locality 
    could lead to the least desirable situation where all the neural network training jobs are scheduled to the only node with GPUs and 
    many of them stay in pending state chronically.
    \item \textit{Contention of Limited Resources.} When the total resource demands of co-located task components of different jobs exceeds 
    the available resources of that node, resource contention happens. Considering that the maximum workloads a node can process 
    mainly depends on the CPU cycle frequency and related hardware performance indexes \cite{choi2004fine,rountree2011practical}, how to 
    allocate resources exhaustively for processing workloads of different job types to reduce the resource contention is of great concern.
    \item \textit{Unknown Arrival Patterns of Jobs.} In real-life scenarios, the workload dispatching decisions should be made online without 
    the knowledge of future job arrivals. The lack of the global information of the problem space could lead to a local optimum.
    \item \textit{Jobs may be Deadline-Sensitive.} Some multi-server jobs have an explicit deadline. For instance, in AI application 
    related companies, a trained deep neural network model is usually guaranteed to be put into service on a particular date. The workload 
    dispatching policy should ensure the training can be finished before deadline even at the cost of performance degeneration. In this 
    case, we can actively reduce the input workload size of data batch to meet the deadline.
\end{itemize}

A majority of workload dispatching and scheduling policies for (multi-server) jobs are proposed by formulating either continuous or 
combinatorial optimization problems with scenario-oriented constraints 
\cite{8917749,gautam2015survey,BSP,8486422,attiya2020job,zhang2020evolving,liang2020data,narayanan2020heterogeneity}. 
The decision variables are either the resource allocation of multiple dimensions or the workload size that each node processes for 
each job. Meanwhile, the optimization target is either job complete \& weighted flow time or the utility that measures the overall 
system efficiency. To solve these optimization programs, algorithms are designed based on various theoretical approaches such as 
relaxed integer programming \cite{BSP}, online primal-dual approaches \cite{8486422}, heuristics \cite{attiya2020job,zhang2020evolving}, 
deep reinforcement learning \cite{liang2020data,narayanan2020heterogeneity}, etc. Many of these policies are effective when facing service 
locality and resource contention even in online settings. However, when adding the deadline-aware constraint, the theoretical 
superiority of these algorithms cannot be guaranteed. Deadline-aware job schedulers are designed mainly from a system approach 
\cite{flow-time,yarn,map-reduce}. For instance, Hu \textit{et al.} present a scheduling framework in big-data platforms, with the 
purpose of minimizing the average workflow turn-around time, to meet job deadlines \cite{flow-time}. A preemption-supported scheduler 
named DAPS is designed for Hadoop YARN clusters \cite{yarn}. In addition, Cheng \textit{et al.} propose a deadline-aware Hadoop job 
scheduler that takes future resource availability into consideration when minimizing job deadline misses \cite{map-reduce}. Nevertheless, 
theoretically guaranteed online workload dispatching policies for deadline-aware multi-server jobs are missing in existing literature. 
Here the theoretical guarantee is that, could we design an $\alpha$-competitive ($\alpha \geq 1$ and the smaller, the better) online 
workload dispatching policy for deadline-aware multi-server jobs, such that the utilities of jobs and the cluster can be maximized 
simultaneously.

To fill the theoretical gap, in this paper, we study a general online workload dispatching problem for deadline-aware multi-server jobs. 
The jobs we consider have a specific size of input workloads, and each of them has an explicit arrival time and deadline to be finished. 
For any computing cluster with heterogeneous and depletable resources on each node, we propose an $\alpha$-competitive policy where 
$\alpha \geq 2$ to decide $(i)$ where to launch each task component instance of the arrived jobs and $(ii)$ the size of workloads that 
each task component processes. The policy is built on the so-called spatio-temporal \textit{resource mesh} of nodes and resource reservation 
is automatically realized. We formulate the problem as a \textit{social welfare maximization} program and solve it online with several well 
designed pseudo functions. The social welfare is defined as the sum of the utilities of jobs and the utility of the computing cluster. 
From the job side, the utility of each job is proportional to the workloads that processed, which is determined by the computation resources 
allocated to it. For instance, for large-scale distributed training, the more data samples (or training epochs) each task component processes, 
the better the trained model. From the cluster side, the utility can be any zero-startup non-decreasing function with diminishing return 
that measures the overall system efficiency or resource fairness. All the max-min, proportional fairness or $\alpha$-fairness are good 
choices \cite{fairness}. We provide rigorous analysis to show that the proposed policy is $\alpha$-competitive for some $\alpha$ at least 
$2$. The theoretical superiority is also validated with simulations and the results show that it distinctly outperforms baselines. Our main 
contributions are summarized as follows.
\begin{enumerate}
    \item We study a general online workload dispatching problem for deadline-aware multi-server jobs from the theoretical perspective. 
    We establish the spatio-temporal resource mesh model and solve the problem with the target of maximizing the social welfare of the 
    system.
    \item We propose an online policy which yields a competitive ratio at least $2$ for general utility settings. Particularly, 
    it has a polynomial complexity when all the utilities are linear and share the same coefficient. The theoretical superiority of 
    the proposed policy is rigorously analyzed and verified with simulations. 
\end{enumerate}
The rest of this paper is organized as follows. We formally introduce the system model and formulate the online workload dispatching 
problem in Sec. \ref{s2}. We present the design details of the online policy with rigorous theoretical analysis in Sec. \ref{s3}. 
We demonstrate the numerical results in Sec. \ref{s4}, and discuss related work in Sec. \ref{s5}. Finally, we conclude this paper 
in Sec. \ref{s6}.

\section{System Model and Problem Formulation}\label{s2}
We consider a computing cluster of heterogeneous VM (and physical) nodes. Let us use $\mathcal{K}$ to denote the set of nodes and index 
each of them by $k$. Key notations used in this paper is summarized in Table \ref{key}.

\begin{table}[htbp]
    \begin{center}
    \caption{\label{key}Summary of key notations.}   
    \begin{tabular}{l|l}    
        \toprule
        {\textsc{Notation}}& {\textsc{Description}}\\[+0.1mm]
        \midrule
        $\mathcal{K}$ & The set of nodes\\[+0.7mm]
        $k \in \mathcal{K}$ & The $k$-th node in $\mathcal{K}$\\[+0.7mm]
        $\mathcal{N}$ & The set of multi-server jobs\\[+0.7mm]
        $n \in \mathcal{N}$ & The $n$-th job in $\mathcal{N}$\\[+0.7mm]
        $a_n, \forall n \in \mathcal{N}$ & The arrival time of job $n$\\[+0.7mm]
        $d_n, \forall n \in \mathcal{N}$ & The strict deadline of job $n$\\[+0.7mm]
        $\rho_n, \forall n \in \mathcal{N}$ & The target workload size of job $n$\\[+0.7mm]
        $\mathcal{K}_n$ & The set of nodes available to job $n$\\[+0.7mm]
        $\mathcal{T}$ & The set of time slots\\[+0.7mm]
        $\tau$ & The length of each time slot in $\mathcal{T}$\\[+0.7mm]
        $\mathcal{R} = \mathcal{K} \times \mathcal{T}$ & The resource mesh\\[+0.7mm]
        $r \in \mathcal{R}$ & The $r$-th resource unit in $\mathcal{R}$\\[+0.7mm]
        $C_r, \forall r \in \mathcal{R}$ & The maximum processable workloads of $r$\\[+0.7mm]
        $\mathcal{R}_n, \forall n \in \mathcal{N}$ & The set of resource units available to $n$\\[+0.7mm]
        $x_{nr}$ & The size of workloads of $n$ dispatched to $r$\\[+0.7mm]
        $\overline{\chi}_{nr}$ & The maximum processing capacity of $r$ for $n$\\[+0.7mm]
        $f_n (\cdot)$ & The utility of job $n$\\[+0.7mm]
        $g (\cdot)$ & The utility of the computing cluster $n$\\[+0.7mm]
        \bottomrule
    \end{tabular}
    \end{center}
\end{table}

\subsection{Spatio-Temporal Resource Mesh}\label{s2.1}
Each node is capable of processing a set of heterogenous multi-server jobs arriving in sequence with different input workload sizes. Let us denote 
the set of jobs as $\mathcal{N}$ and index each of them by $n$. Each job $n$ has a target input workload of size $\varrho_n$ (in MB). 
$\forall n \in \mathcal{N}$, we use $a_n$ and $d_n$ to represent its arrival time and strict deadline to be finished. To maximize the social welfare 
\textit{from a long-term vision}, we consider the time horizon from $\min_{n \in \mathcal{N}} a_n$ to $\max_{n \in \mathcal{N}} d_n$ and evenly divide the 
horizon into slots of length $\tau$. Let us use $\mathcal{T}$ to denote the set of time slots and index each of them with $t$. The time slot length 
$\tau$ can be set as the minimum instance reserved time, for example, 1 hour for AWS spot 
instance\footnote{https://aws.amazon.com/ec2/spot/pricing/}. 

\begin{figure}[htbp]
    \centerline{\includegraphics[width=2.6in]{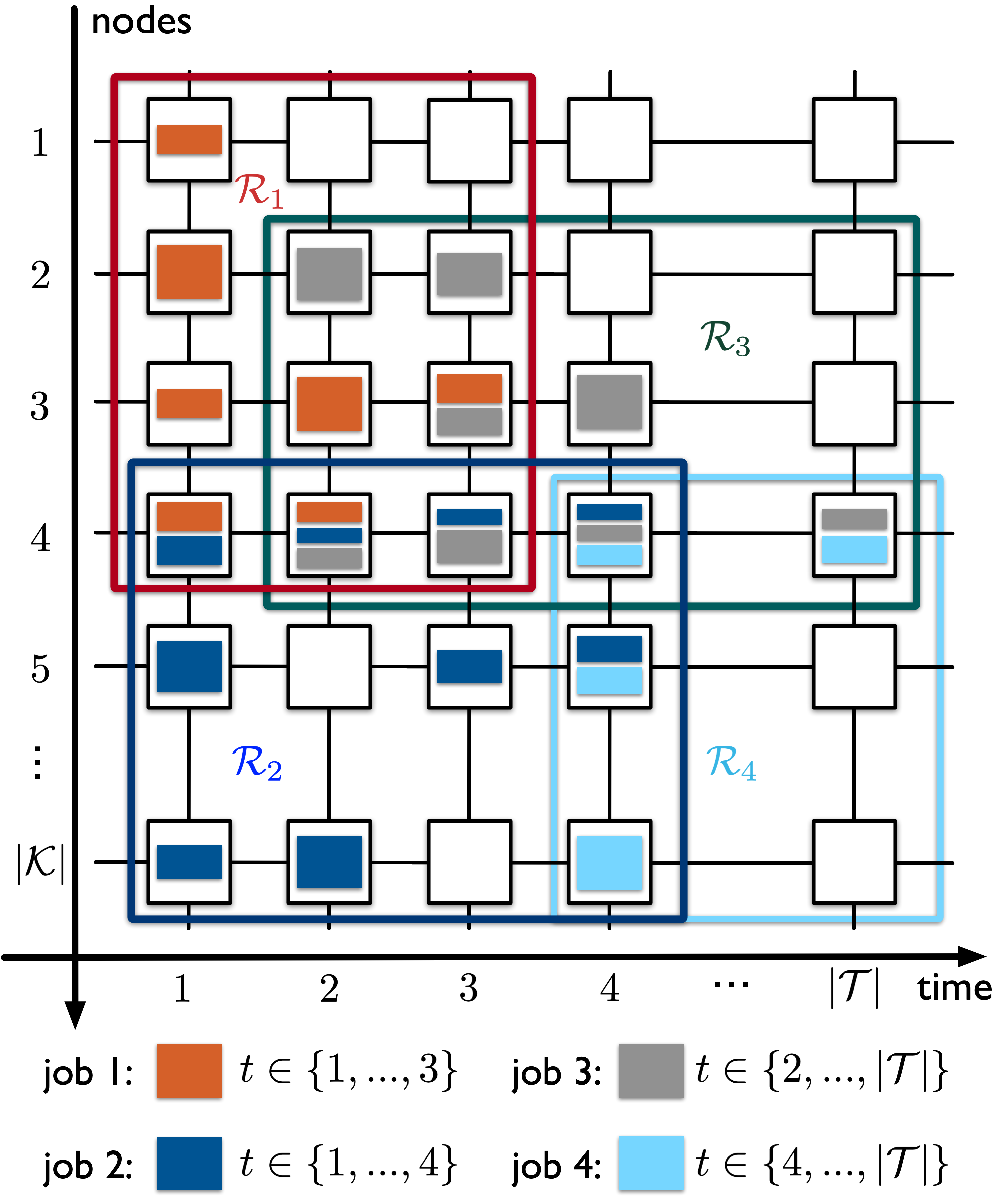}}
    \caption{Available resource units of four exmaple jobs in the resource mesh. Whether resource unit $r$ is available to job $n$ is 
    decided by the service locality constraint and the available time zone of $n$.}
    \label{fig1}
\end{figure}

To manipulate the nodes in $\mathcal{K}$ from both dimension of time and space, we introduce a spatio-temporal resource division model called 
\textit{resource mesh}. We use $\mathcal{R} \triangleq \mathcal{K} \times \mathcal{T}$ to denote the set of resource units and index each of them 
by $r$. Each resource unit $r$, defined as a node in a time slot, can process at most $C_r$ workloads, limited by its hardware performance indexes 
such as the CPU cycle frequency and the GPU clock speed. For distributed model training, $C_r$ indicates the maximum data samples that can be 
processed by $r$ during the given time range. This value could be obtained by a variety of approaches from static code analysis to profiling previous 
runs based on hardware heterogeneity \cite{serverless2}. For each job $n \in \mathcal{N}$, we use
\begin{equation}
    \mathcal{R}_n \triangleq 
    \Big\{r_{kt} \in \mathcal{R} \mid \Big\lceil \frac{a_n}{\tau} \Big\rceil \leq t \leq \Big\lfloor \frac{d_n}{\tau} \Big\rfloor, k \in \mathcal{K}_n \Big\}
\end{equation}
to denote its set of available resource units, where $\mathcal{K}_n \subseteq \mathcal{K}$ is the set of nodes that satisfy the service locality 
of job $n$. Fig. \ref{fig1} gives an example.

\subsection{Utility Functions}\label{s2.2}
For each job $n \in \mathcal{N}$, we need to decide $(i)$ how many task component instances should be initialized, $(ii)$ which resource 
units to place them, and $(iii)$ how many workloads that each task component should process. Our target is to maximize the social welfare, 
i.e., the sum of all jobs' utilities and the utility of the computing cluster. Formally, we use $x_{nr}$ to denote the size of workloads 
dispatched to $r \in \mathcal{R}_n$ and $\overline{\chi}_{nr}$ to denote the maximum workload processing capacity of $r$ for job $n$. 
It results to the constraint $0 \leq x_{nr} \leq \overline{\chi}_{nr}$. Note that $C_r$ is designed to indicate the overall processing 
capacity while $\overline{\chi}_{nr}$ is the workload processing capacity to the specific job $n$ of the $r$-th resource unit. 

We take a zero-startup utility $f_n: [\vec{0}, \overline{\vec{\chi}}_{n}] \to \mathbb{R}$, where 
$\overline{\vec{\chi}}_n \triangleq \big[\overline{\chi}_{nr}\big]_{r \in \mathcal{R}_n}$, as the measurement of satisfaction for job $n$. 
As a widely accepted assumption in previous works \cite{load-balancing1,energy-lb2,zhang2018load,dpos-base,dpos}, we require 
$\{f_n\}_{n\in\mathcal{N}}$ to be non-decreasing, concave, and continuously differentiable on each dimension $r$. Proportional fairness and 
$\alpha$-fairness are good options for $\{f_n\}_{n\in\mathcal{N}}$ \cite{network-utility}. Note that we allow jobs to have different 
utilities. For each job $n \in \mathcal{N}$, its utility is defined as the sum of seperate sub-utilities achieved through each available 
resource unit:
\begin{equation}
    f_n(\vec{x}_n) \triangleq \sum_{r \in \mathcal{R}_n} f_{nr} (x_{nr}), \forall n \in \mathcal{N},
\end{equation}
where $\vec{x}_n \triangleq [x_{nr}]_{r \in \mathcal{R}_n}$. For a given job $n$, $f_{nr}$ can also be different on different 
$r \in \mathcal{R}_n$. To sum up, a multi-server job can be described with the quadruple $\{ \varrho_n, \mathcal{R}_n, \overline{\vec{\chi}}_n, f_n \}$.

To simplify the problem, the utility of the computing cluster 
\begin{equation*}
    g(\vec{x}): [0, x_{nr}]^{\mathcal{N} \times \mathcal{R}} \to \mathbb{R}
\end{equation*}
is defined as the maximized weighted resource utilization efficiency aggregated over each job:
\begin{equation}
    g(\vec{x}) \triangleq \sum_{n \in \mathcal{N}} \sum_{r \in \mathcal{R}_n} \beta_{nr} \cdot \frac{x_{nr}}{C_r}, \forall n \in \mathcal{N}, r \in \mathcal{R}_n,
    \label{g}
\end{equation}
where $\frac{x_{nr}}{C_r}$ is used to indicate the fractional resource consumed by $x_{nr}$, and $\beta_{nr}$ is the weight of provisioning 
total resource of $r$ to $n$. We further use the function $g_{nr} (x_{nr})$ to denote the weighted fractional utility $\beta_{nr} \cdot \frac{x_{nr}}{C_r}$.
In \eqref{g}, $\vec{x}$ is the overall decision variable and $\vec{x} \triangleq [\vec{x}_n]_{n \in \mathcal{N}}$.


\subsection{Online Social Welfare Maximization}\label{s2.3}
Based on the above content, we formulate the social welfare maximization problem as follows:
\begin{flalign}
    \mathcal{P}_1: \max_{\{\vec{x}_n\}_{n \in \mathcal{N}}} 
    &\sum_{n \in \mathcal{N}} f_n (\vec{x}_n) + \sum_{n \in \mathcal{N}} \sum_{r \in \mathcal{R}_n} g_{nr} (x_{nr}) \qquad \qquad \nonumber \\
    s.t. &\sum_{r \in \mathcal{R}_n} x_{nr} \leq \varrho_n, \forall n \in \mathcal{N}, \label{p1_con1} \\
    &x_{nr} = 0, \forall n \in \mathcal{N}, r \in \mathcal{R} \backslash \mathcal{R}_n, \label{p1_con2} \\
    &\sum_{n \in \mathcal{N}} x_{nr} \leq C_r, \forall r \in \mathcal{R}, \label{p1_con3} \\
    &0 \leq x_{nr} \leq \overline{\chi}_{nr}, \forall n \in \mathcal{N}, r \in \mathcal{R}_n. \label{p1_con4}
\end{flalign}
As an offline optimization problem, although $\mathcal{P}_1$ is difficult to solve\footnote{The discrete version of problem $\mathcal{P}_1$ is 
actually a multi-dimensional 0-1 knapsack problem, which is proved to be NP-complete \cite{knapsack-multi}.}, it is built based on complete 
knowledge. However, in online settings, the cluster should not have the information of the $n$-th quadruple 
$\{ \varrho_n, \mathcal{R}_n, \overline{\vec{\chi}}_n, f_n \}$ until job $n$ arrives. To design an efficient 
online workload dispatching policy with the worst-case performance guarantee, we introduce the following notations
\begin{eqnarray}
    \left\{
    \begin{array}{l}
        \iota \triangleq \min_{n \in \mathcal{N}} \min_{r \in \mathcal{R}_n} \Big( \frac{\partial f_n}{\partial x_{nr}} + \frac{\beta_{nr}}{C_r} \Big) \\
        \upsilon \triangleq \max_{n \in \mathcal{N}} \max_{r \in \mathcal{R}_n} \Big( \frac{\partial f_n}{\partial x_{nr}} + \frac{\beta_{nr}}{C_r} \Big).
    \end{array}
    \right.
    \label{lu}
\end{eqnarray}
The ratio between these two constants, i.e., $\frac{\upsilon}{\iota}$, demonstrates the fluctuation of the \textit{marginal} social welfare, 
which will be introduced later. In previous theoretical papers, this ratio is viewed as a known variable and it helps construct the resource 
provision decisions \cite{assume1,ota,assume3,dpos-base,ev-charing1,ota-2}. For example, in \cite{assume3}, the ratio is set as $36$ in default. 

It is worth noting that the online decision $\vec{x}_n$ made for job $n$ when it arrives implies the idea
of \textit{resource reservation}. Support that at time $t$, job $n$ arrives. Formally, for job $n$, by solving $\mathcal{P}_1$ online, 
the resource provision decisions 
\begin{equation*}
    \Big\{x_{nr_{kt'}} \mid r_{kt'} \in \mathcal{R}_n, t' > t \Big\}
\end{equation*}
are the resource reservation results for executing job $n$.

\section{Algorithm Design}\label{s3}
The key challenge to solve $\mathcal{P}_1$ in online settings is that the dispatching of each job's workloads to each resource unit 
are coupled because of \eqref{p1_con3}. Nevertheless, if we could construct several feasible dual variables corresponding to 
$\{\vec{x}_n\}_{n \in \mathcal{N}}$ in $\mathcal{P}_1$, and take these dual variables as the cost for using each resource unit, a near optimal 
solution could be obtained. To implement this, we design several pseudo-social welfare functions with estimated marginal costs. In this design, we 
utilize an important principle for solving online resource provision problems, i.e., \textsl{estimate the cost for processing the workloads of each task component 
as a function of resource surplus} \cite{dpos-base,dpos,assume3,ota,estimate-add}. In the following 
sections, firstly, we show how the pseudo-social welfare functions are constructed. Then, based on these constructed functions, we introduce 
our algorithm \textsc{OnSocMax}. It works by solving several pseudo-social welfare maximization problems polynomially online. To guarantee that \textsc{OnSocMax} is 
$\alpha$-competitive, we analyze the requirements that the marginal cost functions should satisfy. In addition, we give the bound of the 
gap between the competitive ratio achieved by \textsc{OnSocMax} and the optimal competitive ratio of a simplified case under a particular condition. 
In the end, we discuss how to extend \textsc{OnSocMax} to the jobs with non-partitionable workloads and some drawbacks.

\subsection{Pseudo-Social Welfare Function}\label{s3.1}
For each arrived job $n$, we define the pseudo-social welfare function, denoted by $\tilde{\mathcal{W}}_n (\vec{x}_n)$, as
\begin{equation*}
    \Bigg[ f_n(\vec{x}_n) - \sum_{r \in \mathcal{R}_n} \int_{\omega_r^{(n)}}^{\omega_r^{(n)} + x_{nr}} \phi_r(u)du \Bigg] + 
    \sum_{r \in \mathcal{R}_n} g_{nr}(x_{nr}),
\end{equation*}
where $\phi_r$ is a non-decreasing estimation of the marginal cost for the resource unit $r \in \mathcal{R}$ processing unit workload when 
the resource surplus $u \in [0, C_r]$. We define $\phi_r(u) = +\infty$ when $u > C_r$. The non-decreasing property profoundly reflects 
an underlying economic phenomenon, i.e., a thing is valued in proportion to its rarity. The later a job arrives, the higher cost it has 
to pay \cite{dpos-base}. The first component is the pseudo-utility of executing job $n$, which is the utility of it minus the cost to pay. 
The second component is the utility of the computing cluster. If we organize $\tilde{\mathcal{W}}_n(\vec{x}_n)$ as 
\begin{equation*}
    f_n(\vec{x}_n) + \sum_{r \in \mathcal{R}_n} \Bigg[ g_{nr}(x_{nr}) - \int_{\omega_r^{(n)}}^{\omega_r^{(n)} + x_{nr}} \phi_r(u)du \Bigg],
\end{equation*}
the second component can be regarded as the \textit{net profit} of the cluster for processing the workloads of job $n$. In this case, the 
later a job arrives, the harder the resource surplus to meet its requirements before deadline, which results to higher cost. The following 
content applies to both of these two interpretations.

\begin{figure}[htbp]
    \centerline{\includegraphics[width=3.3in]{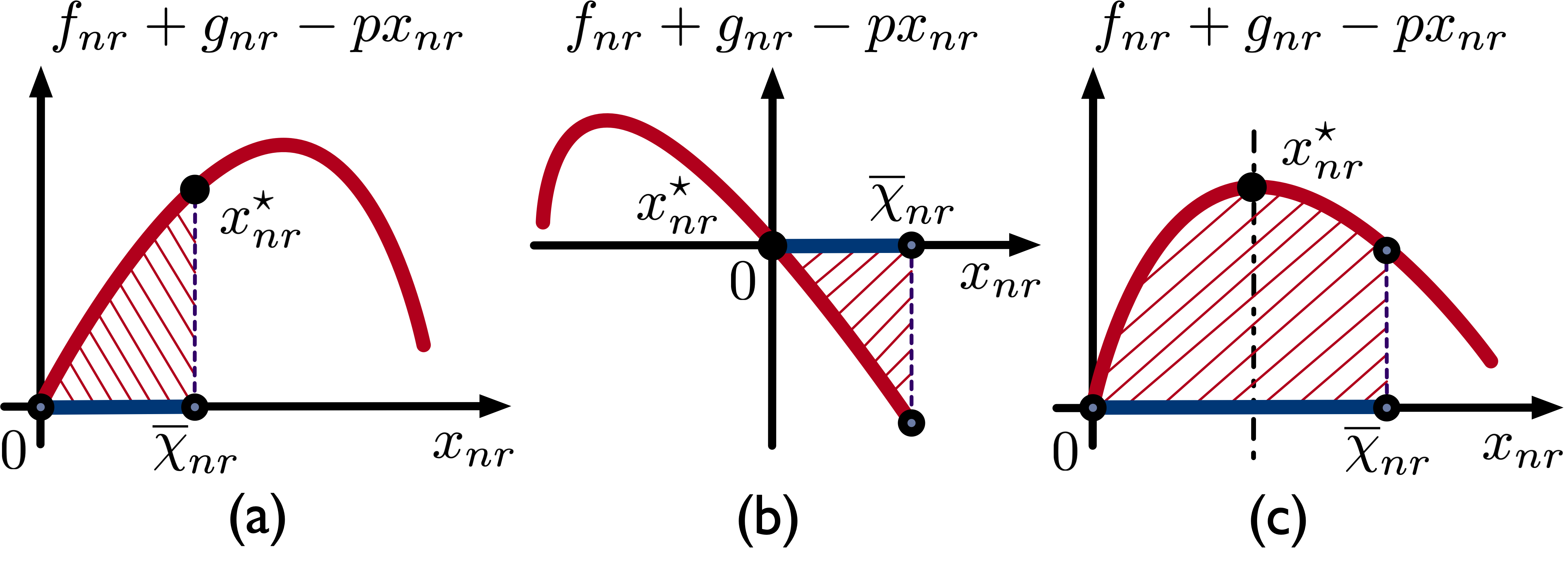}}
    \caption{A visualization on how the three conditions affect the optimal $x_{nr}^*$ of 
    $f_{nr}(x_{nr}) + g_{nr}(x_{nr}) - \big(\phi_r(\omega_r^{(n + 1)}) + \mu_n^*\big) \cdot x_{nr}$, respectively. }
    \label{fig2}
\end{figure}

To bridge connections between the optimal dual variables of $\mathcal{P}_1$ and the optimal solution $\vec{x}_n^*$ that maximizes 
$\tilde{\mathcal{W}}_n$, we firstly introduce the dual problem of $\mathcal{P}_1$ as follows.
\begin{proposition}
    \label{prop1}
    The dual problem of $\mathcal{P}_1$ is:
    \begin{flalign*}
        \mathcal{P}_2: \min_{\vec{\mu}, \vec{\lambda}} \sum_{n \in \mathcal{N}} \sum_{r \in \mathcal{R}} \xi_{nr} (\mu_n + \lambda_r) 
            + \sum_{n \in \mathcal{N}} \mu_n \varrho_n + \sum_{r \in \mathcal{R}} \lambda_r C_r \\
            s.t. \quad \eqref{p1_con2}, \eqref{p1_con4}, \vec{\mu} \geq \vec{0}, \vec{\lambda} \geq \vec{0}, \qquad \qquad \qquad 
    \end{flalign*}
    where
    \begin{equation}
        \xi_{nr} (p) \triangleq \max_{x_{nr} \in [0, \overline{\chi}_{nr}]} \bigg[ f_{nr} (x_{nr}) + \Big( g_{nr} (x_{nr}) - p \cdot x_{nr} \Big) \bigg],
        \label{conjugate}
    \end{equation}
    and $\vec{\mu} \triangleq [\mu_n]_{n \in \mathcal{N}}$ and $\vec{\lambda} \triangleq [\lambda_r]_{r \in \mathcal{R}}$ are the dual 
    variables corresponding to \eqref{p1_con1} and \eqref{p1_con3}, respectively.
\end{proposition}
\begin{proof}
    The result is immediate with Lagrangian.
\end{proof}
Essentially, $\xi_{nr}(\cdot)$ is the convex conjugate of the fractional social welfare $f_{nr} + g_{nr}$.
Taking a closer look at the conjugate $\xi_{nr} (p)$ and the pseudo social welfare $\tilde{\mathcal{W}}_n (\vec{x}_n)$, if we 
could find appropriate $p^*$ and $\vec{x}_n^*$, we can bridge their connection through 
\begin{equation}
    \tilde{\mathcal{W}}_n (\vec{x}_n^*) \approx \sum_{r \in \mathcal{R}_n} \xi_{nr} (p^*).
    \label{approx}
\end{equation}
Based on this, we can interpret $p$ as the marginal cost for processing unit workload \cite{dpos-base}. 
We bridge the subtle connection between $\xi_{nr}$ and $\tilde{\mathcal{W}}_n$ in the following proposition, which is crucial for the design 
of \textsc{OnSocMax}.
\begin{proposition}
    \label{prop2}
    $\forall n \in \mathcal{N}, r \in \mathcal{R}$, when $\phi_r(C_r) \geq \upsilon$, if $(i)$ $\vec{x}_n^* \triangleq [x_{nr}^*]_{r \in \mathcal{R}_n}$ 
    and $\mu_n^*$ are respectively the optimal primal solution and the optimal dual solution to \eqref{p1_con1} of the following problem $\mathcal{P}_3$:
    \begin{flalign*}
        \mathcal{P}_3: \max_{\vec{x}_n} \tilde{\mathcal{W}}_n (\vec{x}_n) \\
        s.t. \quad \eqref{p1_con1}, \eqref{p1_con2}, \eqref{p1_con4},
    \end{flalign*}
    and $(ii)$ the resource usage level $\omega_r$ is updated with 
    \begin{equation}
        \left\{
            \begin{array}{l}
                \omega_r^{(n + 1)} = \omega_r^{(n)} + x_{nr}^* \\
                \omega_r^{(1)} = 0,
            \end{array}
        \right.
        \label{omega}
    \end{equation}
    then, $x_{nr}^*$ is also the optimal solution that maximizes $\xi_{nr}(p)$ given $p = \phi_r(\omega_r^{(n + 1)}) + \mu_n^*$.
\end{proposition}
\begin{proof}
    By the definition of the non-decreasing marginal cost function $\phi_r(\cdot)$, we can find that it is discontinuous at $C_r$. Thus, 
    when $\phi_r(C_r) \geq \upsilon$, there must exist a resource usage level $\overline{\omega}_r \leq C_r$ such that 
    $\phi_r (\overline{\omega}_r) = \upsilon$. Note that the function $f_{n} + \sum_{r \in \mathcal{R}_n}g_{nr}$ is non-decreasing and its 
    derivative on $r$ is not more than $\upsilon$\footnote{This conclusion can be obtained with \eqref{lu}.}. Therefore, when the input of 
    $\phi_r$ is $\omega_r^{(n)} + x_{nr}$, suppose $\omega_r^{(n)} + x_{nr} \leq \overline{\omega}_r$. Consequently, the derivative of the 
    integral function
    \begin{equation}
        \Phi_r (x_{nr}) \triangleq \int_{\omega_r^{(n)}}^{\omega_r^{(n)} + x_{nr}} \phi_r (u) du
        \label{integral}
    \end{equation}
    is continuous, non-decreasing, and convex when $x_{nr} \leq \overline{\omega}_r - \omega_r^{(n)}$. The convexity is because $\Phi_r'$, i.e., 
    $\phi_r$, is non-decreasing. Thus, $\mathcal{P}_3$ is a convex optimization program and its optimal solution can be obtained through KKT 
    conditions. Let us use $x_{nr}^*$, $\mu_n^*$, $\gamma_{nr}^*$, and $\zeta_{nr}^*$ to denote the optimal primal and dual solutions of 
    $\mathcal{P}_3$ ($\mu_n^*$ to \eqref{p1_con1} while $\gamma_{nr}$ and $\zeta_{nr}$ to the right part and left part of \eqref{p1_con4}, 
    respectively). The KKT conditions of $\mathcal{P}_3$ are listed below.
    \begin{eqnarray}
        &\left\{
            \begin{array}{l}
                f_{nr}'(x_{nr}^*) + \frac{\beta_{nr}}{C_r} - \phi_r \big(\omega_r^{(n+1)}\big) - \mu_n^* = \gamma_{nr}^* - \zeta_{nr}^* \\
                \gamma_{nr}^* \big( x_{nr}^* - \overline{\chi}_{nr} \big) = 0 \\
                \zeta_{nr}^* \cdot x_{nr}^* = 0 \\
                \mu_n^* \Big( \sum_{r \in \mathcal{R}_n} x_{nr}^* - \varrho_n \Big) = 0.
            \end{array}
        \right.
        \label{kkt}
    \end{eqnarray}
    With KKT conditions \eqref{kkt}, we show that the optimal solution $x_{nr}^*$ of $\mathcal{P}_3$ simultaneously optimizes the conjugate 
    $\xi_{nr} (p)$ given $p = \phi_r(\omega_r^{(n + 1)}) + \mu_n^*$, i.e., 
    \begin{flalign}
        \xi_{nr} \Big(\phi_r(\omega_r^{(n + 1)}) + \mu_n^*\Big) &= f_{nr}(x_{nr}^*) + g_{nr} (x_{nr}^*) \nonumber \\
        &- \Big( \phi_r(\omega_r^{(n + 1)}) + \mu_n^* \Big) x_{nr}^*.
        \label{connection}
    \end{flalign}

    \underline{\textbf{Case I}}: When $f_{nr}'(x_{nr}^*) + \frac{\beta_{nr}}{C_r} > \phi_r(\omega_r^{(n + 1)}) + \mu_n^*$, 
        $\tilde{\mathcal{W}}_n$ is an increasing function on dimension $r$ under \eqref{p1_con1}. Thus, we have $x_{nr}^* = \overline{\chi}_{nr}$, 
        which leads to 
        \begin{equation}
            f_{nr}'(\overline{\chi}_{nr}) + \frac{\beta_{nr}}{C_r} > \phi_r(\omega_r^{(n + 1)}) + \mu_n^*.
            \label{discuss1}
        \end{equation}
        \eqref{discuss1} indicates that 
        \begin{equation*}
            f_{nr} (x_{nr}) + \Big[ g_{nr} (x_{nr}) - p x_{nr} \Big]
        \end{equation*}
        is monotone increasing in feasible region 
        $[0, \overline{\chi}_{nr}]$ by setting $p$ as $\phi_r(\omega_r^{(n + 1)}) + \mu_n^*$. Therefore, 
        \begin{equation*}
            x_{nr}^* = \overline{\chi}_{nr} = 
        \argmax_{0 \leq x_{nr} \leq \overline{\chi}_{nr}} \bigg[ f_{nr} (x_{nr}) + g_{nr} (x_{nr}) - p \cdot x_{nr} \bigg],
        \end{equation*}
        which means 
        $x_{nr}^*$ maximizes both $\mathcal{P}_3$ and the conjugate $\xi_{nr}(\cdot)$ simultaneously given $p = \phi_r(\omega_r^{(n + 1)}) + \mu_n^*$. Thus, 
        \eqref{connection} holds.

    \underline{\textbf{Case II}}: When $f_{nr}'(x_{nr}^*) + \frac{\beta_{nr}}{C_r} < \phi_r(\omega_r^{(n + 1)}) + \mu_n^*$, similarly, 
        we have $x_{nr}^* = 0$, which leads to 
        \begin{equation}
            f_{nr}'(0) + \frac{\beta_{nr}}{C_r} < \phi_r(\omega_r^{(n)}) + \mu_n^*
            \label{discuss2}
        \end{equation}
        and
        \begin{equation*}
            \omega_r^{(n + 1)} = \omega_r^{(n)} + 0 = \omega_r^{(n)}.
        \end{equation*}
        Analogously, \eqref{discuss2} means that 
        \begin{equation*}
            f_{nr} (x_{nr}) + \Big[ g_{nr} (x_{nr}) - p x_{nr} \Big]
        \end{equation*}
        is monotone decreasing in feasible region $[0, \overline{\chi}_{nr}]$ 
        by setting $p = \phi_r(\omega_r^{(n + 1)}) + \mu_n^*$. Therefore, 
        \begin{equation*}
            x_{nr}^* =0 = 
        \argmax_{0 \leq x_{nr} \leq \overline{\chi}_{nr}} \bigg[ f_{nr} (x_{nr}) + g_{nr} (x_{nr}) - p \cdot x_{nr} \bigg],
        \end{equation*}
        which also leads to \eqref{connection}.

    \underline{\textbf{Case III}}: When $f_{nr}'(x_{nr}^*) + \frac{\beta_{nr}}{C_r} = \phi_r(\omega_r^{(n + 1)}) + \mu_n^*$, $x_{nr}^*$ 
        is an maximum of 
        \begin{equation*}
            f_{nr} (x_{nr}) + \Big[ g_{nr} (x_{nr}) - p x_{nr} \Big]
        \end{equation*}
        given $p = \phi_r(\omega_r^{(n + 1)}) + \mu_n^*$ since 
        $\gamma_{nr}^* = \zeta_{nr}^* = 0$. Thus, 
        \begin{equation*}
            x_{nr}^* \in \argmax_{0 \leq x_{nr} \leq \overline{\chi}_{nr}} \bigg[ f_{nr} (x_{nr}) + g_{nr} (x_{nr}) - p \cdot x_{nr} \bigg],
        \end{equation*}
        which means 
        \eqref{connection} holds.

    All the three conditions are visualized in Fig. \ref{fig2}.
\end{proof}

So far we have analyzed the properties of the pseudo-social welfare functions and the conjugates. In the following sections, we will 
firstly give the design details of the online algorithm \textsc{OnSocMax}. Then, we will illustrate that, to make \textsc{OnSocMax} $\alpha$-competitive 
for some underlying $\alpha$, what requirements the marginal cost functions $\{\phi_{r}\}_{r \in \mathcal{R}}$ should satisfy.

\subsection{\textsc{OnSocMax} Design}\label{s3.2}
\textsc{OnSocMax} is built on solving $\mathcal{P}_3$ for each newly arrived job $n$ in sequence. The procedure is captured in 
\textbf{Algorithm \ref{algo}}. We place a hat on top of variables that denote the variables involved in \textsc{OnSocMax}. 

\begin{figure}
    \removelatexerror
    \begin{algorithm}[H]
        \label{algo}
        \caption{\textsc{OnSocMax}}
        \KwIn{$\{C_r\}_{r \in \mathcal{R}}$ and $\{g_{nr}\}_{n \in \mathcal{N}, r \in \mathcal{R}}$}
        \KwOut{Online solution to $\mathcal{P}_1$ and final utilizations for the resource mesh}
        $\forall r \in \mathcal{R}: \hat{\omega}_r^{(1)} \leftarrow 0$\\
        \While{a new multi-server job $n$ arrives}
        {
            Receive the quadruple $\{ \varrho_n, \mathcal{R}_n, \overline{\vec{\chi}}_n, f_n \}$\\
            Get the (near) optimal solution $\hat{\vec{x}}_{n}$ of $\mathcal{P}_3$\\
            \For(in parallel){$r \in \mathcal{R}_n$}
            {
                $\hat{\omega}_r^{(n+1)} \leftarrow \hat{\omega}_r^{(n)} + \hat{x}_{nr}$ \tcp*[f]{Update utilization}\\
            }
            $n \leftarrow n + 1$\\
        }
        \Return{$\{\hat{\vec{x}}_n\}_{n \in \mathcal{N}}$ and $\big\{\hat{\omega}_r^{(|\mathcal{N}| + 1)}\big\}_{r \in \mathcal{R}}$}
    \end{algorithm}
\end{figure}

Although $\mathcal{P}_3$ is a convex program, we cannot obtain its analytic solution with \eqref{kkt} directly. To solve it iteratively, 
we transform it into the following problem $\mathcal{P}'_3$:

\begin{flalign}
    \mathcal{P}'_3: \min_{\vec{x}_n} \sum_{r \in \mathcal{R}_n} &\bigg[ \Phi_r (x_{nr}) - f_{nr} (x_{nr}) - g_{nr} (x_{nr}) \bigg] \nonumber \\
    s.t. \quad &\sum_{r \in \mathcal{R}_n} x_{nr} - \varrho_n + s = 0, \label{p3-new-cons1}\\
    &x_{nr} - \overline{\chi}_{nr} + l_r = 0, \forall r \in \mathcal{R}_n, \qquad  \label{p3-new-cons2}\\
    &q_r - x_{nr} = 0, \forall r \in \mathcal{R}_n, \label{p3-new-cons3}\\
    &s, l_r, q_r \geq 0, \forall r \in \mathcal{R}_n, \label{p3-new-cons4}
\end{flalign}
where $\vec{l} \triangleq [l_r]_{r \in \mathcal{R}_n}$, $\vec{q} \triangleq [q_r]_{r \in \mathcal{R}_n}$, and $s$ are introduced 
slack variables. In addition, we define the dual variables respectively to \eqref{p3-new-cons1}, \eqref{p3-new-cons2}, and \eqref{p3-new-cons3} 
as $\mu$, $\vec{y} \triangleq [ y_r ]_{r \in \mathcal{R}_n}$, and $\vec{z} \triangleq [z_r]_{r \in \mathcal{R}_n}$. 

The augmented Lagrangian of $\mathcal{P}'_3$ is
\begin{flalign}
    & \qquad \qquad \qquad L_{\sigma} (\vec{x}_n, s, \vec{l}, \vec{q}, \mu, \vec{y}, \vec{z}) \nonumber \\
    &= \sum_{r \in \mathcal{R}_n} \bigg[ \Phi_r (x_{nr}) - f_{nr} (x_{nr}) - g_{nr} (x_{nr}) \bigg] \nonumber \\
    &+ \mu \Big( \sum_{r \in \mathcal{R}_n} x_{nr} - \varrho_n + s \Big) + \sum_{r \in \mathcal{R}_n} y_r \Big( x_{nr} - \overline{\chi}_{nr} + l_r \Big) \nonumber \\
    &+ \sum_{r \in \mathcal{R}_n} z_r (q_r - x_{nr}) + \frac{\sigma}{2} \rho (\vec{x}_n, s, \vec{l}, \vec{q}),
    \label{aug}
\end{flalign}
where the penalty function $\rho (\cdot)$ is defined as
\begin{flalign}
    \rho (\vec{x}_n, s, \vec{l}, \vec{q}) &\triangleq \sum_{r \in \mathcal{R}_n} \Big( q_r - x_{nr} \Big)^2 + 
    \Big( \sum_{r \in \mathcal{R}_n} x_{nr} - \varrho_n + s \Big)^2 \nonumber \\
    &+ \sum_{r \in \mathcal{R}_n} \Big( x_{nr} - \overline{\chi}_{nr} + l_r \Big)^2,
\end{flalign}
and $\sigma > 0$ is the penalty coefficient. If we consider the augmented Lagrangian $L_{\sigma} (\vec{x}_n, s, \vec{l}, \vec{q}, \mu, \vec{y}, \vec{z})$ as the 
function of the slack variables $[s, \vec{l}, \vec{q}]$, to minimize it, we can get their optimal values:
\begin{eqnarray}
    \left\{
        \begin{array}{l}
            s^* = \max \Big\{ - \frac{\mu}{\sigma} + \varrho_n - \sum_{r \in \mathcal{R}_n} x_{nr}, 0 \Big\}, \\
            l^*_r = \max \Big\{ - \frac{y_r}{\sigma} + \overline{\chi}_{nr} - x_{nr}, 0 \Big\}, \forall r \in \mathcal{R}_n, \\
            q_r^* = \max \Big\{ - \frac{z_r}{\sigma} + x_{nr}, 0 \Big\}, \forall r \in \mathcal{R}_n.
        \end{array}
    \right.
    \label{slq}
\end{eqnarray}
Taking \eqref{slq} into \eqref{aug}, we have 
\begin{flalign}
    L_{\sigma} ( &\vec{x}_n, \mu, \vec{y}, \vec{z}) = \sum_{r \in \mathcal{R}_n} \Big[ \Phi_r (x_{nr}) - f_{nr} (x_{nr}) - g_{nr} (x_{nr}) \Big] \nonumber \\
    &+ \frac{\sigma}{2} \Bigg[ \max \Big\{ \frac{\mu}{\sigma} + \sum_{r \in \mathcal{R}_n} x_{nr} - \varrho_n, 0 \Big\}^2 - \frac{\mu^2}{\sigma^2} \Bigg] \nonumber \\
    &+ \frac{\sigma}{2} \sum_{r \in \mathcal{R}_n} \Bigg[ \max \Big\{ \frac{y_r}{\sigma} + x_{nr} -  \overline{\chi}_{nr}, 0 \Big\}^2 - \frac{y_r^2}{\sigma^2} \Bigg] \nonumber \\
    &+ \frac{\sigma}{2} \sum_{r \in \mathcal{R}_n} \Bigg[ \max \Big\{ \frac{z_r}{\sigma} - x_{nr}, 0 \Big\}^2 - \frac{z_r^2}{\sigma^2} \Bigg].
    \label{aug2}
\end{flalign}
Besides, we define the constraint violation degree $v (\cdot)$ of a solution $\vec{x}_n$ given $(\mu, \vec{y}, \vec{z})$ by 
\begin{flalign}
    v (\vec{x}_n \mid \mu, \vec{y}, \vec{z}) &\triangleq \max \bigg\{ \sum_{r \in \mathcal{R}_n} x_{nr} - \varrho_n, - \frac{\mu}{\sigma_\kappa} \bigg\} \nonumber \\
    &+\sum_{r \in \mathcal{R}_n} \max \bigg\{ x_{nr} - \overline{\chi}_{nr}, - \frac{y_r}{\sigma_\kappa} \bigg\} \nonumber \\
    &+ \sum_{r \in \mathcal{R}_n} \max \bigg\{ -x_{nr}, - \frac{z_r}{\sigma_\kappa} \bigg\}.
    \label{vio}
\end{flalign}
Based on \eqref{aug2}, we can solve $\mathcal{P}'_3$ approximatively with the augmented Lagrangian method. The procedure is summarized in 
\textbf{Algorithm \ref{algo2}}, which is used to substitute step 4 of \textsc{OnSocMax}. 

\begin{figure}
    \removelatexerror
    \begin{algorithm}[H]
        \label{algo2}
        \caption{The augmented Lagrangian method}
        \KwIn{$\{C_r\}_{r \in \mathcal{R}}$, $\{g_{nr}\}_{n \in \mathcal{N}, r \in \mathcal{R}}$, $\{ \varrho_n, \mathcal{R}_n, \overline{\vec{\chi}}_n, f_n \}$, and 
        $\{ \omega_r^{(n)}\}_{r \in \mathcal{R}_n}$}
        \KwOut{The (near) optimal solution $\hat{\vec{x}}_{n}$ of $\mathcal{P}_3$}
        Initialize the primal and dual variables $\vec{x}_n^0, \mu^0, \vec{y}^0, \vec{z}^0$\\
        Initialize the penalty coefficient $\sigma_0 > 0$ and
        \begin{flalign*}
            0 < \theta_1 \leq \theta_2 \leq 1, p > 1
        \end{flalign*}\\
        Initialize the constraint violation coefficient $\eta_0 = \frac{1}{\sigma_0}$, the precision coefficient $\varepsilon_0 = \frac{1}{\sigma_0^{\theta_1}}$, 
        and their final value $\eta, \varepsilon$\\
        $\kappa \leftarrow 0$\\
        \While{true}
        {
            Get a solution $\vec{x}_n^{\kappa+1}$ of $\min_{\vec{x}_n^\kappa} L_{\sigma_k} (\vec{x}_n^\kappa, \mu^\kappa, \vec{y}^\kappa, \vec{z}^\kappa)$ 
            by gradient descent method which satifies the following precision:
            \begin{flalign*}
                \Big\| \nabla_{\vec{x}_n}  L_{\sigma_k} (\vec{x}_n, \mu^\kappa, \vec{y}^\kappa, \vec{z}^\kappa) \Big\| \leq \eta^\kappa
            \end{flalign*}\\
            Calculate the constraint violation degree $v (\vec{x}_n^{\kappa+1} \mid \mu^\kappa, \vec{y}^\kappa, \vec{z}^\kappa)$ by \eqref{vio}\\
            \uIf{$v (\vec{x}_n^{\kappa+1} \mid \mu^\kappa, \vec{y}^\kappa, \vec{z}^\kappa) \leq \varepsilon_\kappa$}
            {
                \If{$\| \nabla_{\vec{x}_n}  L_{\sigma_k} (\vec{x}_n, \mu^\kappa, \vec{y}^\kappa, \vec{z}^\kappa)\| \leq \eta$ and $v (\vec{x}_n^{\kappa+1} \mid \mu^\kappa, \vec{y}^\kappa, \vec{z}^\kappa) \leq \varepsilon$}
                {
                    \textbf{return} $\vec{x}_n^{\kappa+1}$\\
                }
                \tcc{Update the dual variables}
                $\mu^{\kappa+1} \leftarrow \max \Big\{ \mu^\kappa + \sigma_\kappa \Big( \sum_{r \in \mathcal{R}_n} - \varrho_n \Big), 0 \Big\}$\\
                \For(in parallel){$r \in \mathcal{R}_n$}
                {
                    $y_r^{\kappa+1} \leftarrow \max \Big\{ y_r^\kappa + \sigma_\kappa \Big( x_{nr}^{\kappa+1} - \overline{\chi}_{nr} \Big), 0 \Big\}$\\
                    $z_r^{\kappa+1} \leftarrow \max \Big\{ z_r^\kappa - \sigma_\kappa x_{nr}^{\kappa+1}, 0 \Big\}$\\
                }
                $\sigma_{\kappa+1} \leftarrow \sigma_\kappa$\\
                $\eta_{\kappa+1} = \eta_\kappa / \sigma_{\kappa+1}, \varepsilon_{\kappa+1} \leftarrow \varepsilon_\kappa / \sigma_{\kappa + 1}^{\theta_2}$\\
            }
            \Else
            {
                \tcc{Keep the dual variables unchanged}
                $\mu^{\kappa+1} \leftarrow \mu^\kappa, \vec{y}^{\kappa + 1} \leftarrow \vec{y}^\kappa, \vec{z}^{\kappa+1} \leftarrow \vec{z}^\kappa $\\
                $\sigma_{\kappa+1} \leftarrow p \sigma_\kappa$\\
                $\eta_{\kappa+1} = 1 / \sigma_{\kappa+1}, \varepsilon_{\kappa+1} \leftarrow 1 / \sigma_{\kappa + 1}^{\theta_1}$\\
            }
            $\kappa \leftarrow \kappa + 1$\\
        }
        \Return{the solution of the final iteration $\vec{x}_n^\kappa$}
    \end{algorithm}
\end{figure}

\subsection{Competitive Analysis}\label{s3.add}

\textsc{OnSocMax} is at most polynomial because $\mathcal{P}_3$ is convex and can be solved efficiently in polynomial time with augmented Lagrangian method.
Obviously, $\{\hat{\vec{x}}_n\}_{n \in \mathcal{N}}$ is feasible to $\mathcal{P}_1$. To quantify how “good” \textsc{OnSocMax} is, we adopt 
the standard competitive analysis framework \cite{competitive}. 
\begin{definition}
    For any arrival instance $\mathcal{A}$ of all the multi-server jobs, the competitive ratio for an online algorithm is defined as 
    \begin{equation}
        \alpha \triangleq \max_{\forall \mathcal{A}} \frac{\Theta_{\mathcal{P}_1}^* (\mathcal{A})}{\Theta_{\textrm{\textit{on}}} (\mathcal{A})},
    \end{equation}
    where $\Theta_{\mathcal{P}_1}^* (\mathcal{A})$ is the maximum objective value of $\mathcal{P}_1$, $\Theta_{\textrm{\textit{on}}} (\mathcal{A})$ 
    is the objective function value of $\mathcal{P}_1$ obtained by this online algorithm. 
    \label{def1}
\end{definition}
The competitive ratio quantifies the worst-case ratio between the optimum and the objective obtained by the online algorithm.
The smaller $\alpha$ is, the better the online algorithm. An online algorithm is called $\alpha$-\textit{competitive} 
if its ratio is upper bounded by $\alpha$. 

Now we give the requirements the marginal cost functions $\{\hat{\phi}_{r}\}_{r \in \mathcal{R}}$ should satisfy to guarantee that \textsc{OnSocMax} is 
$\alpha$-competitive for some $\alpha$.
\begin{theorem}
    \label{theo1}
    \textsc{OnSocMax} is $\alpha$-competitive for some $\alpha \geq 1$ if $\forall r \in \mathcal{R}$, the 
    marginal cost function $\hat{\phi}_r$ is in the form of
    \begin{eqnarray}
        \hat{\phi}_r(\omega) = 
        \left\{
            \begin{array}{ll}
                \iota & \omega \in [0, \hat{\varpi}_r) \\
                \hat{\varphi}_r(\omega) & \omega \in [\hat{\varpi}_r, C_r] \\
                +\infty & \omega \in (C_r, +\infty),
            \end{array}
        \right.
        \label{competitive-1}
    \end{eqnarray}
    where $\hat{\varpi}_r$ is a resource utilization threshold, and $\hat{\varphi}_r$ is a non-decreasing function that satisfies 
    \begin{eqnarray}
        \left\{
            \begin{array}{l}
                \hat{\varphi}_r(\omega) C_r \leq \alpha \int_0^\omega \hat{\phi}_r(u)du - \iota \cdot \omega, \quad \omega \in [\hat{\varpi}_r, C_r] \\
                \hat{\varphi}_r(\hat{\varpi}_r) = \iota, \hat{\varphi}_r(C_r) \geq \upsilon.
            \end{array}
        \right.
        \label{competitive-2}
    \end{eqnarray}
\end{theorem}
\begin{proof}
    To prove this result, we refer to the technique named \textit{instance-dependent online primal-dual approach}, proposed in \cite{assume3}. 
    The key idea is to construct a dual solution to $\mathcal{P}_2$ based on the solution $\{\hat{\vec{x}}_n\}_{n \in \mathcal{N}}$ produced by 
    \textsc{OnSocMax}. Then, it uses this dual objective to build the upper bound of the optimum of $\mathcal{P}_1$ with weak duality. When building the upper bound, this technique studies \textit{the worst-case arrival instances} under different scenarios.

    Let us use $\mathcal{B} \triangleq \{\mathcal{A}_1, \mathcal{A}_2, ...\}$ to denote the set of arrival instances of the multi-server jobs, 
    and use $\Theta_{\mathcal{P}_2} (\mathcal{A})$ to denote a feasible objective value of the dual problem $\mathcal{P}_2$ for any arrival instance 
    $\mathcal{A}$. Hereinafter, we just replace $\hat{\omega}_r^{(|\mathcal{N}| + 1)}$ by $\hat{\omega}_r^N$ for simplification. 
    We divide $\mathcal{B}$ into three disjoint sets:
    \begin{eqnarray}
        \left\{
            \begin{array}{l}
                \mathcal{B}_1 \triangleq \{\mathcal{A} \mid 0 \leq \hat{\omega}_r^N < \hat{\varpi}_r, \forall r \in \mathcal{R} \} \\
                \mathcal{B}_2 \triangleq \{\mathcal{A} \mid \hat{\varpi}_r \leq \hat{\omega}_r^N \leq C_r, \forall r \in \mathcal{R} \} \\
                \mathcal{B}_3 \triangleq \mathcal{B} \backslash \big( \mathcal{B}_1 \cup \mathcal{B}_2 \big).
            \end{array}
        \right.
    \end{eqnarray}
    $\mathcal{B}_1$ and $\mathcal{B}_2$ contain the instances whose final utilizations of all resource units in the mesh are below and above 
    the threshold $\hat{\varpi}_r$, respectively. Our goal is to prove that, under the conditions \eqref{competitive-1} and \eqref{competitive-2}, 
    $\forall \mathcal{A} \in \mathcal{B}_1, \mathcal{B}_2, \mathcal{B}_3$ respectively, the following relations hold:
    \begin{equation}
        \alpha \cdot \Theta_{\textrm{\textit{on}}} (\mathcal{A}) \geq \Theta_{\mathcal{P}_2} (\mathcal{A}) \geq \Theta_{\mathcal{P}_1}^* (\mathcal{A}).
    \end{equation}
    In the following analysis, we just drop the parentheses and $\mathcal{A}$ for simplification.

    \underline{\textbf{Case I}}: $\forall \mathcal{A} \in \mathcal{B}_1$, from \eqref{competitive-1} we can find that the marginal costs 
    experienced by all jobs are the same, i.e., $\iota$. In this case, each job $n$ is processed with maximum permitted workloads 
    $\overline{\chi}_{nr}$ on $r \in \mathcal{R}_n$. Thus, $\Theta_{\mathcal{P}_1}^* / \Theta_{\textrm{\textit{on}}} = 1 \leq \alpha$. 
    
    \underline{\textbf{Case II}}: $\forall \mathcal{A} \in \mathcal{B}_2$, we construct a feasible dual solution to $\mathcal{P}_2$ as 
    \begin{eqnarray}
        \left\{
            \begin{array}{ll}
                \hat{\mu}_n = \mu_n^*, &\forall n \in \mathcal{N} \\
                \hat{\lambda}_r = \hat{\phi}_r (\hat{\omega}_r^N) &\forall r \in \mathcal{R},
            \end{array}
        \right.
        \label{omega_2}
    \end{eqnarray}
    where $\mu_n^*$ is the optimal dual solution to $\mathcal{P}_3$ introduced by \eqref{kkt}. Let $p \geq p' \geq 0$ and denote the 
    optimal solution that maximizes the conjugate $\xi_{nr} (p)$ by $\tilde{x}_{nr}$ given $p$. Then, 
    \begin{eqnarray}
        \xi_{nr} (p) &=& f_{nr} (\tilde{x}_{nr}) + \big( g_{nr} (\tilde{x}_{nr}) - p \cdot \tilde{x}_{nr} \big) \nonumber\\
        &\leq& f_{nr} (\tilde{x}_{nr}) + \big( g_{nr} (\tilde{x}_{nr}) - p' \cdot \tilde{x}_{nr} \big) \nonumber\\
        &\leq& \max_{x_{nr}} \Big[ f_{nr} (x_{nr}) + \big( g_{nr} (x_{nr}) - p' \cdot x_{nr} \big) \Big] \nonumber\\
        &=& \xi_{nr} (p'),
        \label{non-increasing}
    \end{eqnarray}
    which indicates that the conjugate $\xi_{nr}(p)$ is non-increasing with $p$. The above derivation uses the fact that $f_{nr} + g_{nr}$ is 
    non-decreasing. Based on weak duality and the non-increasing property of the conjugate, we have
    \begin{flalign*}
        \quad \Theta_{\mathcal{P}_1}^* \leq& \sum_{n \in \mathcal{N}} \sum_{r \in \mathcal{R}} \xi_{nr} \Big(\mu_n^* + \hat{\phi}_r \big(\hat{\omega}_r^N\big)\Big) + \sum_{n \in \mathcal{N}} \mu_n^* \varrho_n  \\
        +& \sum_{r \in \mathcal{R}} \hat{\phi}_r \big(\hat{\omega}_r^N\big) C_r \qquad \qquad \triangleright \text{\textit{the right-side is }} \Theta_{\mathcal{P}_2} \\
        \leq& \sum_{n \in \mathcal{N}} \sum_{r \in \mathcal{R}} \xi_{nr} \Big(\mu_n^* + \hat{\phi}_r \big(\hat{\omega}_r^{(n+1)}\big)\Big) + \sum_{n \in \mathcal{N}} \mu_n^* \varrho_n  \\
        +& \sum_{r \in \mathcal{R}} \hat{\phi}_r \big(\hat{\omega}_r^N\big) C_r \qquad \qquad \triangleright \eqref{non-increasing} \\
        =& \sum_{r \in \mathcal{R}} \bigg[ \hat{\phi}_r \big(\hat{\omega}_r^N\big) C_r - \sum_{n \in \mathcal{N}} \hat{\phi}_r \big(\hat{\omega}_r^{(n+1)}\big) \hat{x}_{nr} \bigg] \\
        +& \sum_{n \in \mathcal{N}} \sum_{r \in \mathcal{R}} \Big( f_{nr}(\hat{x}_{nr}) + g_{nr}(\hat{x}_{nr}) \Big) \triangleq \Theta_{\textrm{\textit{tmp}}}. \triangleright \eqref{connection}
    \end{flalign*}
    The last equality holds because $\hat{x}_{nr}$ simultaneously maximizes $\mathcal{P}_3$ and the conjugate $\xi_{nr}\big(\mu_n^* + \hat{\phi}_r (\hat{\omega}_r^N)\big)$ 
    (result of \textbf{Proposition \ref{prop2}}). Since $\{\hat{\phi}_r\}_{r \in \mathcal{R}}$ are non-decreasing, $\forall n \in \mathcal{N}, r \in \mathcal{R}$, 
    \begin{equation}
        \hat{\phi}_r \big(\hat{\omega}_r^{(n+1)}\big) \hat{x}_{nr} \geq \int_{\hat{\omega}_r^{(n)}}^{\hat{\omega}_r^{(n+1)}} \hat{\phi}_r (u)du.
        \label{28}
    \end{equation}
    \eqref{28} is illustrated in Fig. \ref{fig3}. Further, we have
    \begin{equation}
        \sum_{n \in \mathcal{N}} \hat{\phi}_r \big(\hat{\omega}_r^{(n+1)}\big) \hat{x}_{nr} \geq \int_{\hat{\omega}_r^{(1)}}^{\hat{\omega}_r^N} \hat{\phi}_r (u)du,
        \label{29}
    \end{equation}
    where $\hat{\omega}_r^{(1)} = 0$ because of \eqref{omega}.
    Besides, from Fig. \ref{fig2} we can find that $\xi_{nr} (\mu_n^* + \hat{\phi}_r (\hat{\omega}_r^{(n+1)})) \geq 0$ holds for all the serverless functions.
    Thus, based on \eqref{29}, we have
    \begin{flalign}
        \sum_{n \in \mathcal{N}} \sum_{r \in \mathcal{R}} \Big( f_{nr}(\hat{x}_{nr}) + g_{nr}(\hat{x}_{nr}) \Big) 
        \geq \int_{0}^{\hat{\omega}_r^N} \hat{\phi}_r (u)du.
        \label{30}
    \end{flalign}
    Based on the above results \eqref{29} and \eqref{30}, we have
    \begin{flalign*}
        \Theta_{\textrm{\textit{tmp}}} \leq& \sum_{r \in \mathcal{R}} \bigg[ \hat{\phi}_r \big(\hat{\omega}_r^N\big) C_r - \int_{0}^{\hat{\omega}_r^N} \hat{\phi}_r (u)du \bigg] \\
        +& \sum_{n \in \mathcal{N}} \sum_{r \in \mathcal{R}} \bigg( f_{nr}(\hat{x}_{nr}) + g_{nr}(\hat{x}_{nr}) \bigg) \qquad \triangleright \eqref{29} \\
        <& \sum_{r \in \mathcal{R}} \big(\alpha - 1\big) \int_{0}^{\hat{\omega}_r^N} \hat{\phi}_r (u)du \quad \triangleright \eqref{competitive-2} \text{ \&\textit{ drop }$\iota \cdot \hat{\omega}_r^N$}\\
        +& \sum_{n \in \mathcal{N}} \sum_{r \in \mathcal{R}} \bigg( f_{nr}(\hat{x}_{nr}) + g_{nr}(\hat{x}_{nr}) \bigg) \\
        \leq& \sum_{n \in \mathcal{N}} \sum_{r \in \mathcal{R}} \bigg( f_{nr}(\hat{x}_{nr}) + g_{nr}(\hat{x}_{nr}) \bigg) \alpha. \quad \triangleright \eqref{30}
    \end{flalign*}
    The final experission is exactly $\alpha \cdot \Theta_{\textrm{\textit{on}}}$. Thus, $\Theta_{\mathcal{P}_1}^* / \Theta_{\textrm{\textit{on}}} < \alpha$. 
    
    \begin{figure}[htbp]
        \centerline{\includegraphics[width=2.3in]{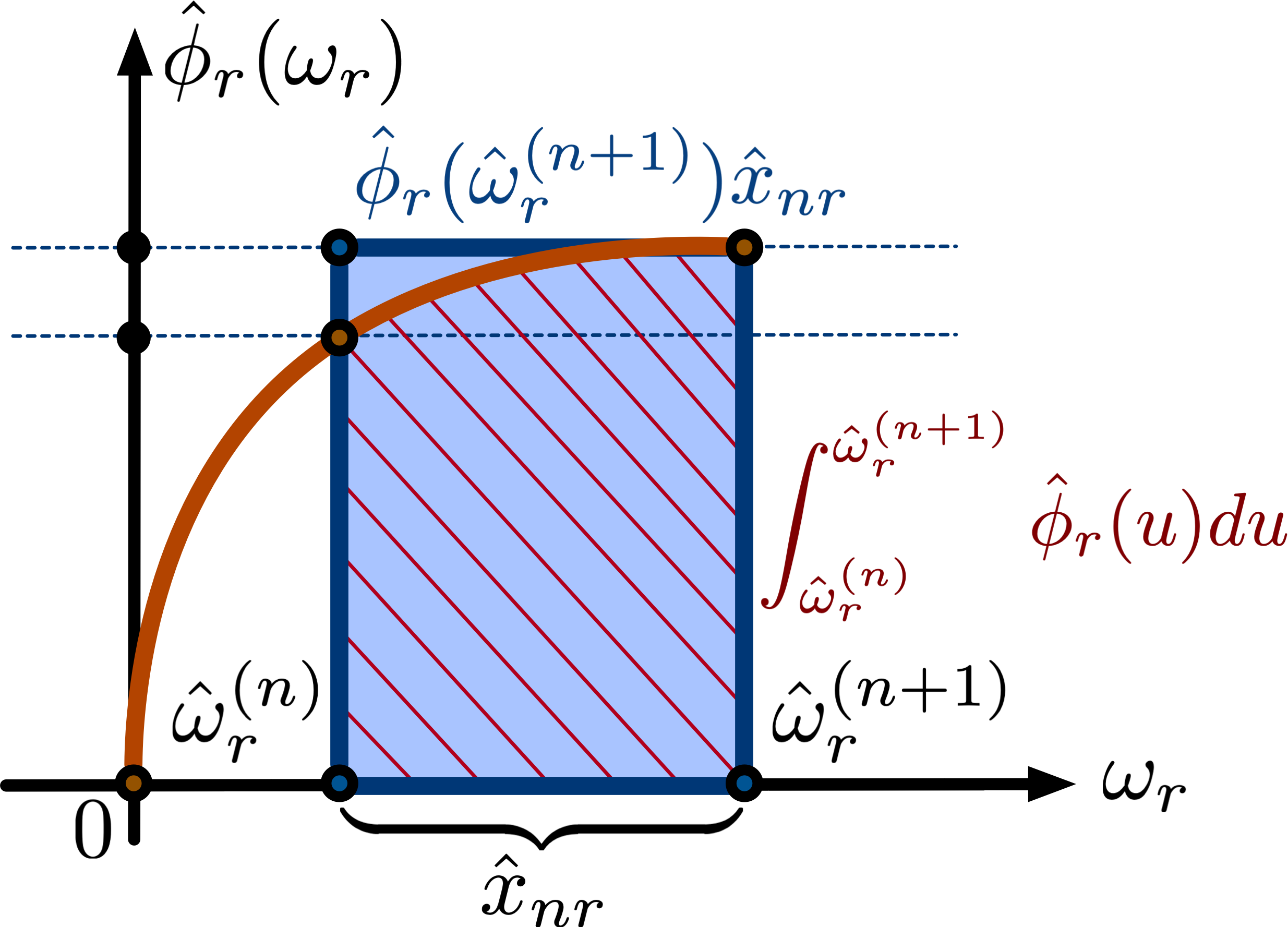}}
        \caption{A visualization of \eqref{28}. The area of the blue rectangle is not less than the area of the shaded region because of the non-decreasing 
        property of $\hat{\phi}_r$. }
        \label{fig3}
    \end{figure}

    \underline{\textbf{Case III}}: $\forall \mathcal{A} \in \mathcal{B}_3$, we define two disjoint sets to split the resource mesh $\mathcal{R}$:
    \begin{eqnarray}
        \left\{
            \begin{array}{l}
                \mathcal{R}_1 \triangleq \{ r \in \mathcal{R} \mid 0 \leq \hat{\omega}_r^N < \hat{\varpi}_r, \forall r \in \mathcal{R} \} \\
                \mathcal{R}_2 \triangleq \{ r \in \mathcal{R} \mid \hat{\varpi}_r \leq \hat{\omega}_r^N \leq C_r, \forall r \in \mathcal{R} \}.
            \end{array}
        \right.
    \end{eqnarray}
    For resource unit $r$ in different sets, the corresponding dual variables are constructed in different ways. We extend $\mathcal{P}_1$ to $\mathcal{P}'_1$ by adding 
    the following constraint:
    \begin{equation}
        \sum_{n \in \mathcal{N}} \sum_{r \in \mathcal{R}_1} x_{nr} \leq \sum_{r \in \mathcal{R}} \hat{\omega}_r^N.
        \label{new_cons}
    \end{equation}
    Apparently, $\mathcal{P}'_1$ is the same as $\mathcal{P}_1$ for \textsc{OnSocMax} since \eqref{new_cons} is not violated by $\{\hat{\vec{x}}_n\}_{n \in \mathcal{N}}$. The dual problem 
    $\mathcal{P}'_2$ to $\mathcal{P}'_1$ is 
    \begin{subequations}
        \begin{flalign*}
            \mathcal{P}'_2: \min_{\vec{\mu}, \vec{\lambda}} \sum_{n \in \mathcal{N}} 
            \Big[& \sum_{r \in \mathcal{R}_1} \xi_{nr} (\mu_n + \lambda_r + \delta) + \sum_{r \in \mathcal{R}_2} \xi_{nr} (\mu_n + \lambda_r) \Big] \\
            +& \sum_{n \in \mathcal{N}} \mu_n \varrho_n + \sum_{r \in \mathcal{R}} \lambda_r C_r + \delta \sum_{r \in \mathcal{R}} \hat{\omega}_r^N
        \end{flalign*}
        \begin{equation*}
            s.t. \quad \eqref{p1_con2}, \eqref{p1_con4}, \vec{\mu} \geq \vec{0}, \vec{\lambda} \geq \vec{0}, \delta \geq 0,
        \end{equation*}
    \end{subequations}
    where $\delta$ is the dual variable corresponding to the newly added constraint \eqref{new_cons}. Then, we construct the dual solution to $\mathcal{P}'_2$ as
    \begin{eqnarray}
        \left\{
            \begin{array}{l}
                \hat{\lambda}_r = 
                \left\{
                    \begin{array}{ll}
                        0 &\forall r \in \mathcal{R}_1\\
                        \hat{\phi}_r (\omega_r^N) &\forall r \in \mathcal{R}_2
                    \end{array}
                \right.\\
                \delta = \iota \\
                \hat{\mu}_n = \mu^*_n \quad \forall n \in \mathcal{N}.
            \end{array}
        \right.
        \label{case3-dual}
    \end{eqnarray}
    Based on \eqref{case3-dual}, we can follow a similar approach as show in Case II to obtain that $\Theta_{\mathcal{P}_1}^* / \Theta_{\textrm{\textit{on}}} \leq \alpha$. 
    A slight difference is that, in Case III, when applying \eqref{competitive-2} to $\Theta'_{\textrm{\textit{tmp}}}$, the result is 
    tightly bounded.
\end{proof}

\textbf{Theorem \ref{theo1}} extends the Two-Point Boundary Value ODEs for designing the marginal cost functions from standard 0-1 knapsack problem to multi-dimensional fractional problems.
Based on \textbf{Theorem \ref{theo1}} and Gronwall's Inequality \cite{gronwall}, we have the detailed design of $\{\hat{\phi}_r\}_{r \in \mathcal{R}}$, which is irrelevant 
with the utilities $\{f_n\}_{n \in \mathcal{N}}$ and $\{g_{nr}\}_{n \in \mathcal{N}, r \in \mathcal{R}}$, as follows.
\begin{theorem}
    \label{theo2}
    For any resource unit $r \in \mathcal{R}$, if the marginal cost function $\hat{\phi}_r$ introduced in the pseudo-social welfare function is designed as 
    \begin{eqnarray*}
        \hat{\phi}_r(\omega) = 
        \left\{
            \begin{array}{ll}
                \iota & \omega \in [0, \hat{\varpi_r}) \\
                \frac{\upsilon - \iota}{\exp (\hat{\alpha}) - \exp ( \frac{\hat{\alpha}}{\hat{\alpha} - 1} )} e^{\big(\frac{\hat{\alpha}}{C_r}\omega\big)} + \frac{\iota}{\hat{\alpha}} & \omega \in [\hat{\varpi_r}, C_r] \\
                +\infty & \omega \in (C_r, +\infty),
            \end{array}
        \right.
    \end{eqnarray*}
    where $\hat{\varpi}_r = \frac{C_r}{\hat{\alpha} - 1}$, then $(i)$ \textsc{OnSocMax} is $\hat{\alpha}$-competitive, where $\hat{\alpha}$ is the solution of 
    \begin{equation}
        \hat{\alpha} - 1 = \frac{1}{\hat{\alpha} - 1} + \ln \frac{\hat{\alpha} \frac{\upsilon}{\iota} - 1}{\hat{\alpha} - 1}
        \label{alpha-circ}
    \end{equation}
    and $(ii)$ when $\hat{\alpha} \geq \frac{\iota}{\upsilon} + 1$, the gap between $\hat{\alpha}$ and the optimal 
    competitive ratio when $|\mathcal{R}| = 1$ is at least $\frac{2}{\sqrt{5} + 1} - \ln \frac{\sqrt{5} + 1}{2} \approx 0.1368$.
\end{theorem}
\begin{proof}
    We firstly introduce the Gronwall's inequality \cite{gronwall} as follows. \textit{$\forall x \in [\underline{x}, \overline{x}]$, if 
    $f(x) \leq a(x) + b(x) \int_{\underline{x}}^x f(u)du$, then 
    \begin{equation}
        f(x) \leq a(x) + b(x) \int_{\underline{x}}^x a(u) \bigg[ \int_u^x b(w)dw \bigg] du,
        \label{gron}
    \end{equation}
    where $f(x)$ is continuous, $a(x)$ and $b(x)$ are integrable and $\forall x \in [\underline{x}, \overline{x}], b(x) \geq 0$. The result remains valid if all the `$\leq$' are 
    replaced by `$=$'.} Applying \eqref{gron} to \eqref{competitive-2} leads to 
    \begin{equation}
        \iota \leq \hat{\varphi}_r(C_r) \leq \frac{\iota}{\hat{\alpha}} + \Bigg[ \frac{\iota \hat{\varpi}_r (\hat{\alpha} - 1)}{C_r} - \frac{\iota}{\hat{\alpha}} \Bigg] e^{\big( \hat{\alpha} \frac{C_r - \hat{\varpi}_r}{C_r} \big)}.
        \label{res2}
    \end{equation}
    Thus, the minimum $\hat{\alpha}$ is achieved when all inequalities in \eqref{competitive-2} and \eqref{res2} are binding, which leads to the design of $\{\hat{\phi}_r\}_{r \in \mathcal{R}}$ 
    and the competitive ratio achieved by \eqref{alpha-circ}.

    In the following, we prove the results of $(ii)$. When $\mathcal{R} = 1$, $\mathcal{P}_1$ degenerates to the general one-way trading (GOT) problem \cite{got}. The optimal 
    competitive ratio is proved to be $1 + \ln (\frac{\upsilon}{\iota})$\cite{got,ota,ota-2,dpos-base,assume3}. With $\hat{\alpha} \geq 1, \frac{\upsilon}{\iota} \geq 1$, 
    let us take $y \geq 1$ as a substitute for $\hat{\alpha} - 1$. Then 
    \begin{flalign*}
        \hat{\alpha} - 1 - \ln \Big( \frac{\upsilon}{\iota} \Big) &= y - \ln \Big( \frac{\upsilon}{\iota} \Big) \qquad \triangleright \text{\textit{with \eqref{alpha-circ}}} \\
        &= \ln \Big[ y + 1 - \frac{\iota}{\upsilon} \Big] + \frac{1}{y} - \ln y \\
        &\triangleq \textsc{Gap} (y).
    \end{flalign*}
    Applying $\ln (x) \leq x - 1, \forall x \geq 1$ to the logarithm in $\textsc{Gap}(y)$, we have $\textsc{Gap}(y) \leq y + \frac{1}{y} - \ln y - \frac{\iota}{\upsilon}$
    given $y \geq \frac{\iota}{\upsilon}$. By analyzing the upper bound of $\textsc{Gap}(y)$, we can easily find that when $y^* = \frac{\sqrt{5} + 1}{2}$, its upper bound 
    is at least $\frac{1}{y^*} - \ln y^*$, which directly leads to the result in $(ii)$.
\end{proof}
By the design of $\hat{\phi}_r(\cdot)$, we observe that $\alpha \geq 2$ holds because $\frac{\upsilon}{\iota} \geq 1$. 
When $\{f_{nr}\}_{n \in \mathcal{N}, r \in \mathcal{R}_n}$ are linear and share the same coefficient, $\hat{\alpha} = 2$. 

\subsection{Extending to Non-Partitionable Workloads}\label{s3.3}
\textsc{OnSocMax} can be easily applied to general jobs whose workloads are not permitted to be partitioned. Specifically, in this case, 
\eqref{p1_con4} is replaced by 
\begin{equation}
    x_{nr} \in \{0, \overline{\chi}_{nr}\}, \forall n \in \mathcal{N}, r \in \mathcal{R}_n,
    \label{p1_cons1_changed}
\end{equation}
and $\overline{\chi}_{nr} = \varrho_n$. To solve the new problem in online settings, we can approximate the marginal cost defined in \eqref{integral} 
with $\hat{\phi}_r (\hat{\omega}_r^{(n)} + \hat{\mu}_{nr}) \hat{x}_{nr}$. With this substitution, the Case III in Fig. \ref{fig2} is 
merged into Case I or Case II, and \textsc{OnSocMax} achieves the same competitive ratio as shown in \textbf{Theorem \ref{theo2}}. 
This approach is exactly the implementation of \eqref{approx}.

\subsection{Discussion on Drawbacks}\label{s3.5}
So far we have demonstrated the design details of \textsc{OnSocMax}. It works by solving several well designed pseudo social welfare functions. 
The problem we formulated is essentially an NP-hard online multi-dimensional knapsack problem. Although \textsc{OnSocMax} is $\alpha$-competitive 
and easy to implement in real-life systems, there are some defects that cannot be ignored and will be in-depth studied in future.

\begin{itemize}
    \item \textit{Communication costs are not analytically counted.} For each multi-server job, the more task component instances we launched, the more 
    communication costs. It can lead to the diminishing returns on the job's utility. Although the utility we considered is allowed to be concave, the 
    elaborate relation between the communication cost and the user satisfaction is not analytically analyzed.
    \item \textit{$C_r$ and $\overline{\chi}_{nr}$ might be hard to determined.} In the established model, $C_r$ is the maximum processable workloads and $\overline{\chi}_{nr}$ is the workload processing capability to job $n$ on the resource unit $r$. Testing tools built on statistical code analysis and profiling techniques 
    might be required for determining these constants. 
\end{itemize}


\section{Simulation Results}\label{s4}
In this section, we conduct several simulations to validate the theoretical superiority of \textsc{OnSocMax}. The experiments are not meant to be 
exhaustive, rather they are used to illustrate the potential of \textsc{OnSocMax}. 

\subsection{Experimental Setup}\label{s4.1}
\textit{Jobs and Nodes.} We consider a cluster with $10$ nodes in the time horizon of $24$ time slots. The processing capacity 
of nodes are generated from an \textit{i.i.d.} Gaussian $\mathcal{N}(\mu = 20, \sigma = 2)$. By setting $\tau$ as $60$ minutes, 
the time horizon represents one day. We set the number of multi-server jobs as $20$. The number of job arrivals in each time slot 
follows a Poisson distribution with a mean of $2.03$ requests, which is independent of other time slots in this day. The deadline of each job 
is calculated based on the arrive time and the maximum service duration of it, where the latter is generated by an Exponential distribution\footnote{Note that the Poisson distribution and the exponential distribution are used for data generation, which is not necessary to \textsc{OnSocMax}.} with a mean of 
$4$ time slots (2 hours). Each job has an input workload whose size is generated from a Normal distribution $\mathcal{N}(\mu = 18, \sigma = 3)$. The 
workload processing capability of each job on each node is generated from a Normal distribution $\mathcal{N}(\mu = 7, \sigma = 1)$. 

\textit{Utilities and Parameters.} $\forall n \in \mathcal{N}$, the utility of job $n$ is set as a zero-startup, non-decreasing concave function. 
We study $f_{nr}$ in three cases: linear, logarithmic, and polynomial. Specifically, for each $n \in \mathcal{N}$, $r \in \mathcal{R}_n$, 
\begin{flalign*}
    f_{nr}(x) = \left\{
        \begin{array}{ll}
            ax & \textrm{\textit{linear}} \\
            a \log(x + 1) & \textrm{\textit{log}} \\
            a \sqrt{x} & \textrm{\textit{poly}},
        \end{array}
    \right.
\end{flalign*}
where the coefficient $a$ is generated from a uniform distribution in $[1, 3]$. Similarly, the parameter $\beta_{nr}$ in $g_{nr}(\cdot)$ is 
generated from the uniform distribution in $[0.1, 0.5]$. 

\textit{Hyper-Parameters of \textsc{OnSocMax}.} There are many algorithmic hyper-parameters involve setting in \textbf{Algorithm \ref{algo2}}. For example, 
the initial penalty coefficient $\sigma_0$, the initial constraint violation coefficient $\eta_0$, the initial precision coefficient $\varepsilon_0$, etc. 
Their default settings are listed in Table \ref{tab2}. In the last line, learning rate and decay are parameters involved in step 6 of \textbf{Algorithm \ref{algo2}}.

\begin{table}[htbp]   
    \begin{center}
    \caption{\label{tab2}Default hyper-parameter settings.}   
    \begin{tabular}{c|c|c|c}    
        \toprule
        {\textsc{Parameter}} & {\textsc{Value}} & {\textsc{Parameter}} & {\textsc{Value}}\\[+0.1mm]
        \midrule
        $\sigma_0$ & $0.97$ & $p$ & $1.002$\\[+0.7mm]
        $\theta_1$ & $0.99$ & $\theta_2$ & $0.999$ \\[+0.7mm]
        $\eta$ & $0.1$ & $\varepsilon$ & $10$ \\[+0.7mm]
        learning rate & $2$ & decay & $0.95$ \\[+0.7mm]
        \bottomrule   
    \end{tabular}  
    \end{center}
\end{table}

\begin{figure*} 
    \centering 
    \subfigure[Social welfare under linear utilities.]{ 
      \includegraphics[height=1.33in]{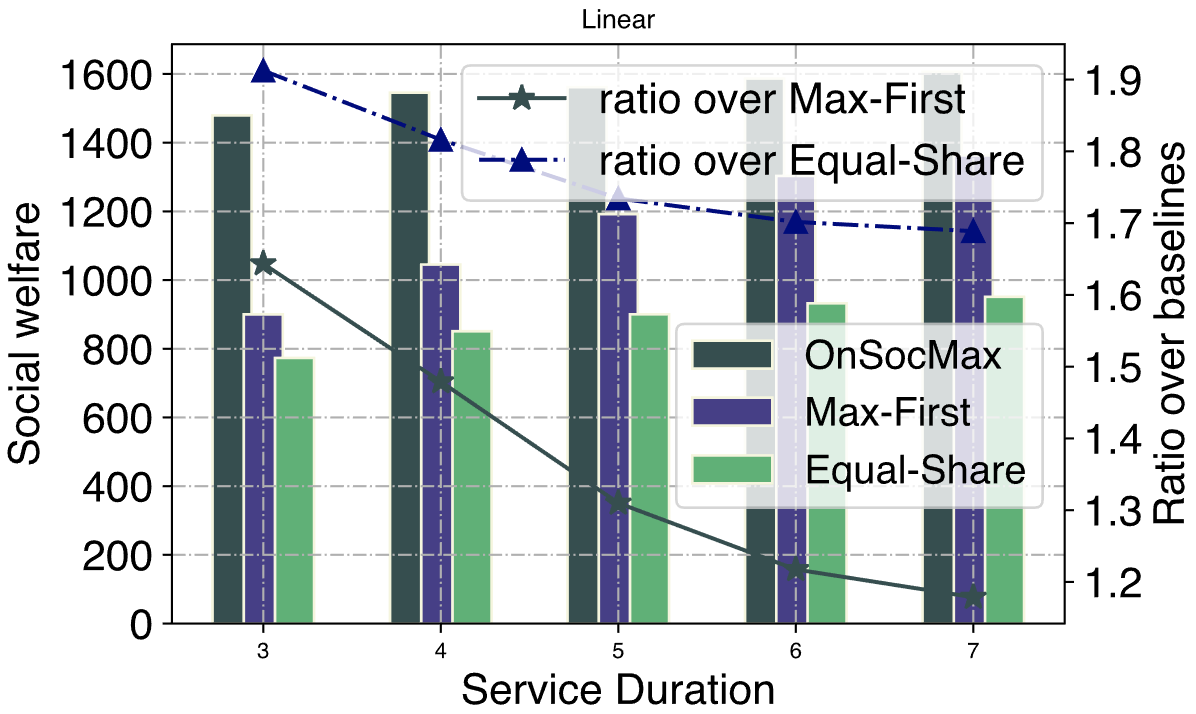}
    }
    \subfigure[Social welfare under poly utilities.]{ 
      \includegraphics[height=1.33in]{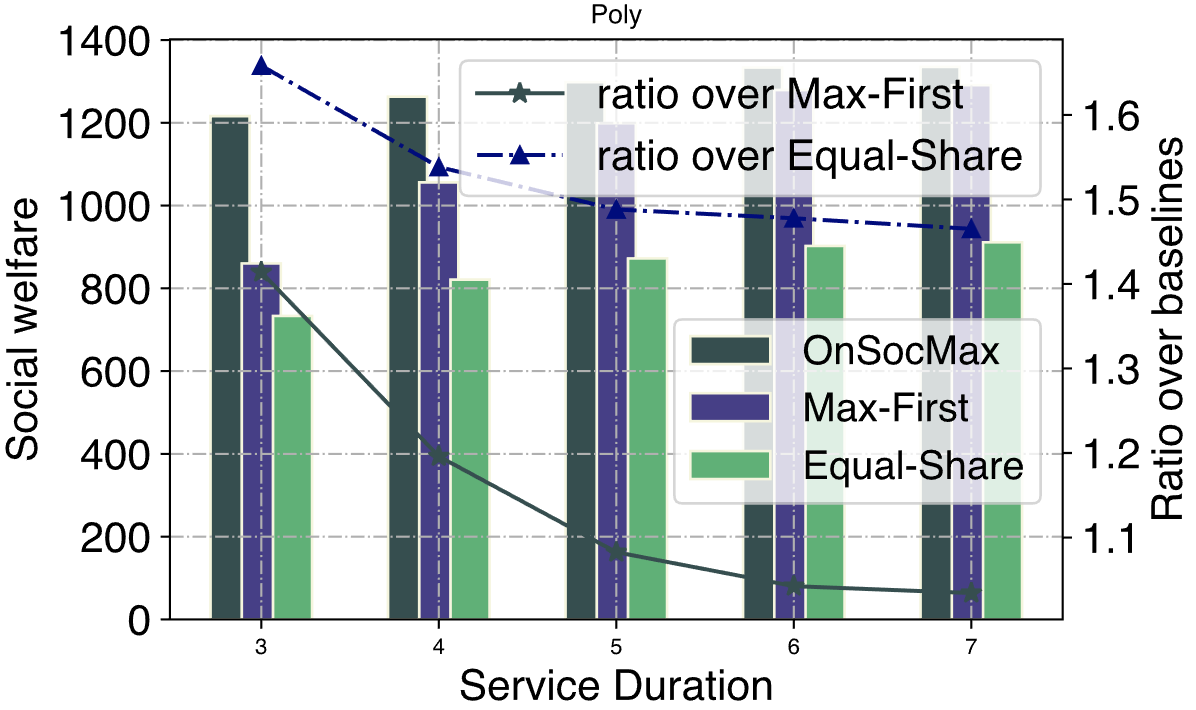}
    }
    \subfigure[Social welfare under log utilities.]{ 
      \includegraphics[height=1.33in]{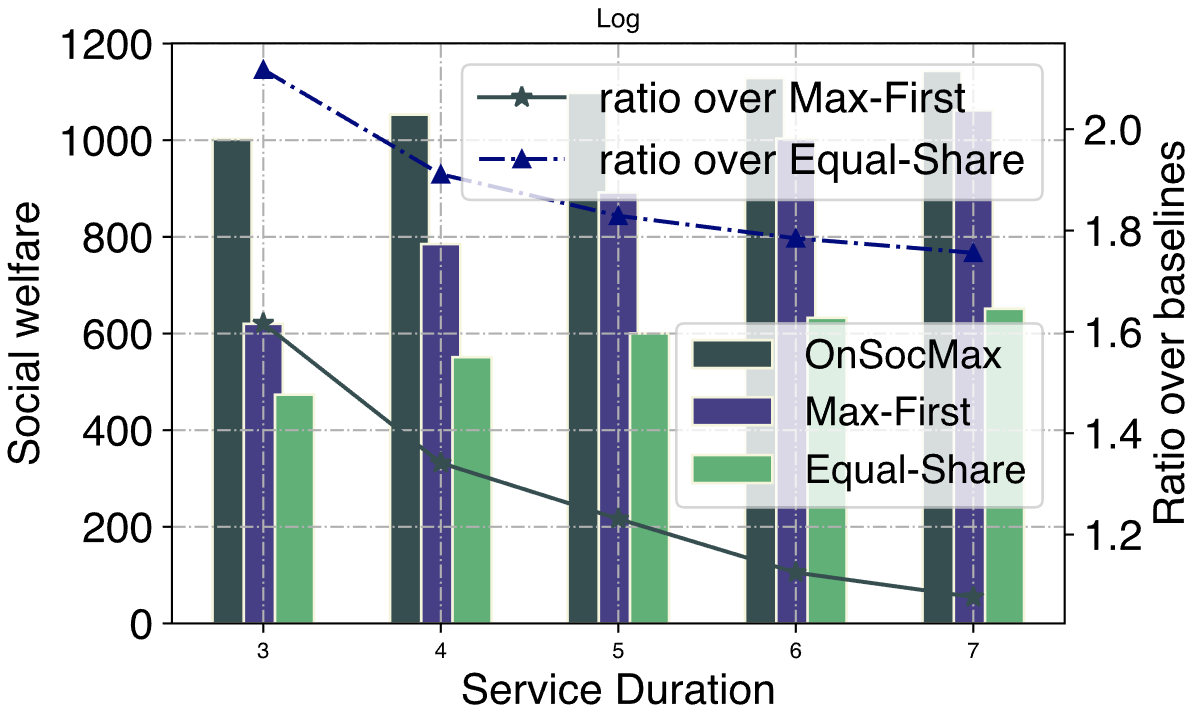}
    }
    \caption{Social welfare of three algorithms under different service duration settings.} 
    \label{exp1}
\end{figure*}

\begin{figure*} 
    \centering 
    \subfigure[Social welfare under linear utilities.]{ 
      \includegraphics[height=1.33in]{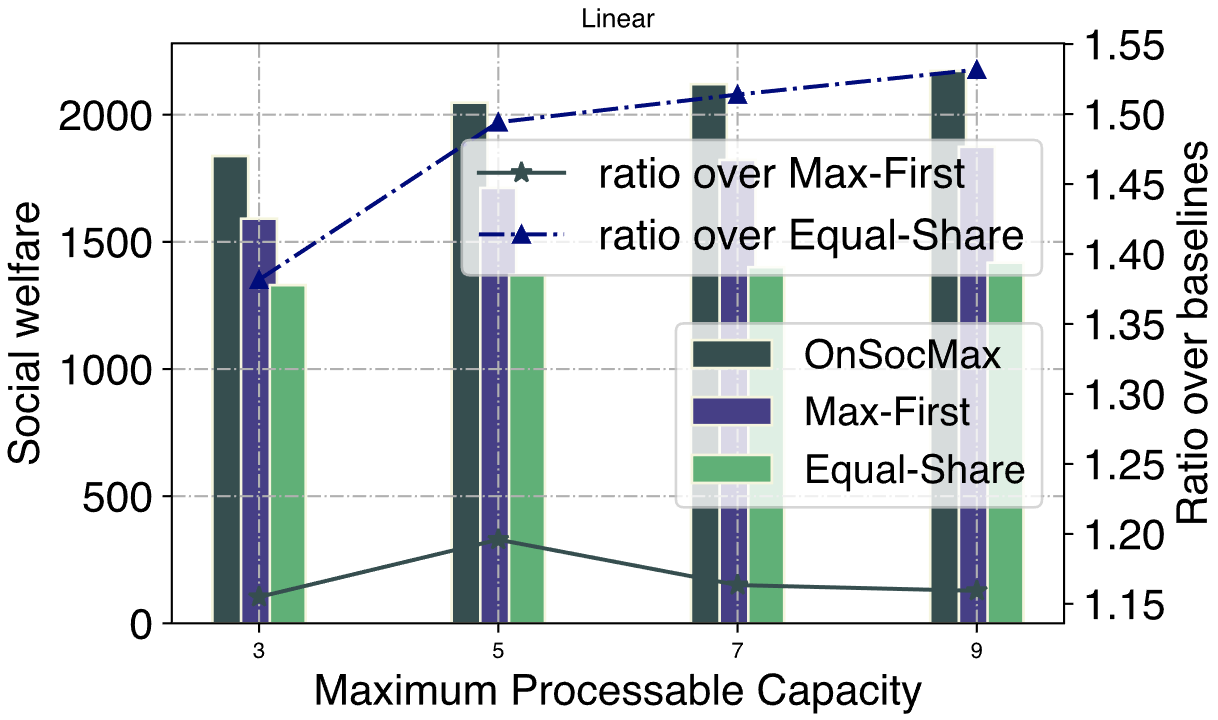}
    }
    \subfigure[Social welfare under poly utilities.]{ 
      \includegraphics[height=1.33in]{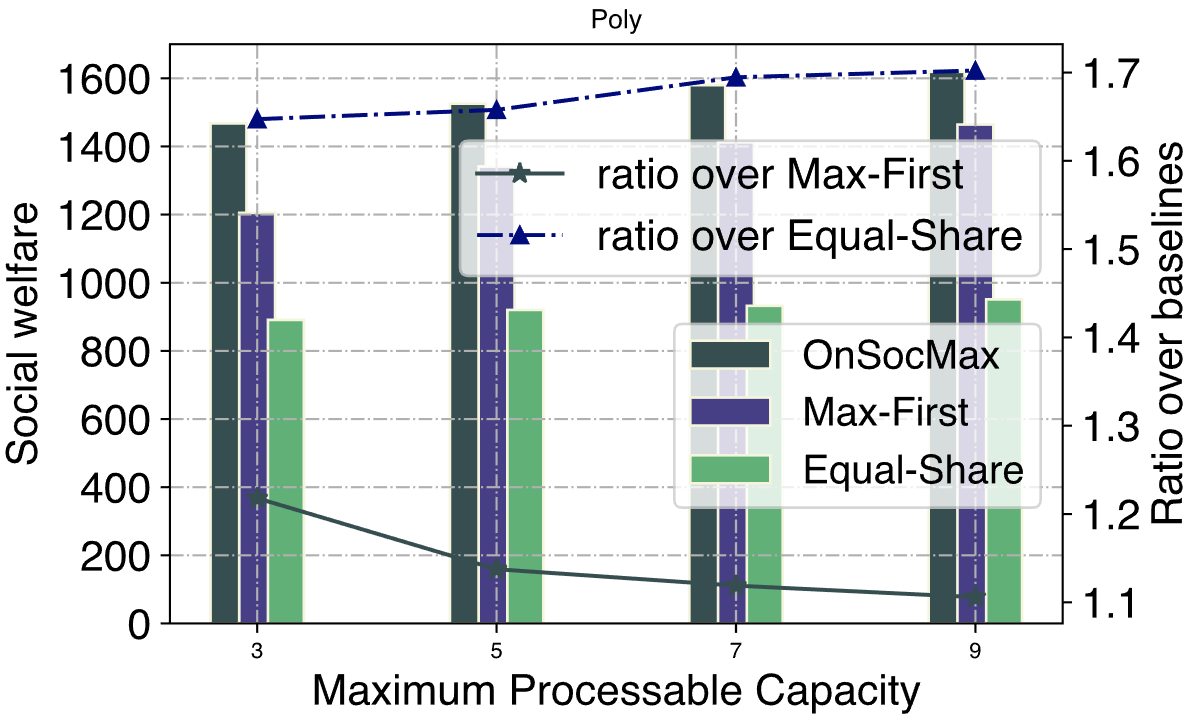}
    }
    \subfigure[Social welfare under log utilities.]{ 
      \includegraphics[height=1.33in]{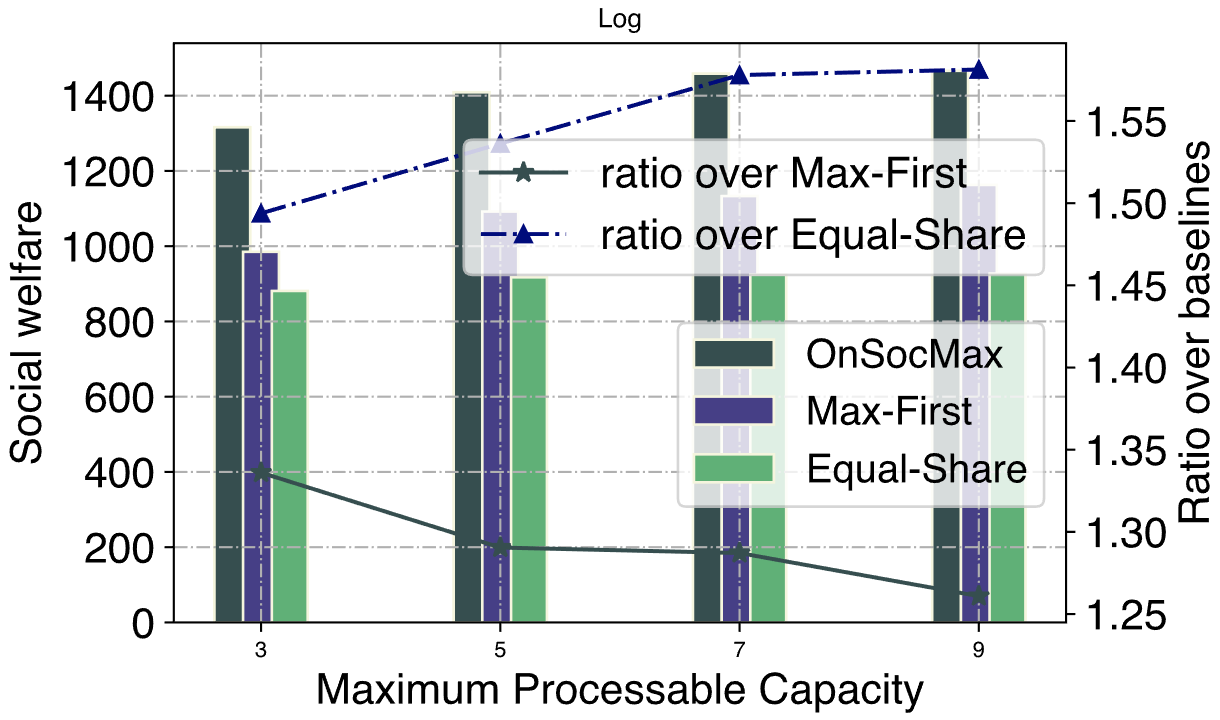}
    }
    \caption{Social welfare of three algorithms under different workload size settings.} 
    \label{exp2}
\end{figure*}

\textit{Algorithms Compared.} We compare \textsc{OnSocMax} with two handcrafted online algorithms. 
\begin{itemize}
    \item \textsc{Max-First}. In \textsc{Max-First}, each node always serves the job with the highest \textit{myopic social welfare}, i.e., 
    the sum of the utility of the chosen job and the utility for serving it in each time slot is maximized. \textsc{Max-First} is myopic because 
    it always maximizes the \textit{partial} social welfare that it sees.
    \item \textsc{Equal-Share}. In \textsc{Equal-Share}, a node serves every newly arrived job with equal opportunity within its capacity limit.
\end{itemize}

\subsection{Simulation Results}\label{s4.2}
In the following content, firstly, we verify the performance of \textsc{OnSocMax} against the handcrafted policies on the social welfare achieved. 
Then, we analyze the robustness of \textsc{OnSocMax} under different parameter settings. 

\textit{Theoretical Superiority.} 
Fig. \ref{exp1} and Fig. \ref{exp2} show the social welfare achieved under different service duration and input workloads settings of jobs. 
We can observe that all the algorithms achieve higher social welfare when the two variables increase. The reason is 
that, when the capacity of nodes are sufficient, increasing the service duration and the maximum input workloads of jobs can increase the opportunities 
of being fully served. Nevertheless, \textsc{OnSocMax} always performs the best among these online algorithms. 
Besides, \textsc{OnSocMax} performs the best for linear job utilities. This is because the logarithmic and polynomial utilities have diminishing returns, which could 
increase the fluctuation ratio $\frac{\upsilon}{\iota}$. This will cause more jobs be served with their marginal costs fall into the second 
segment of $\hat{\phi}_r(\omega)$, which further leads to the decrease of social welfare.  

\begin{figure}[htbp]
    \centerline{\includegraphics[width=3.3in]{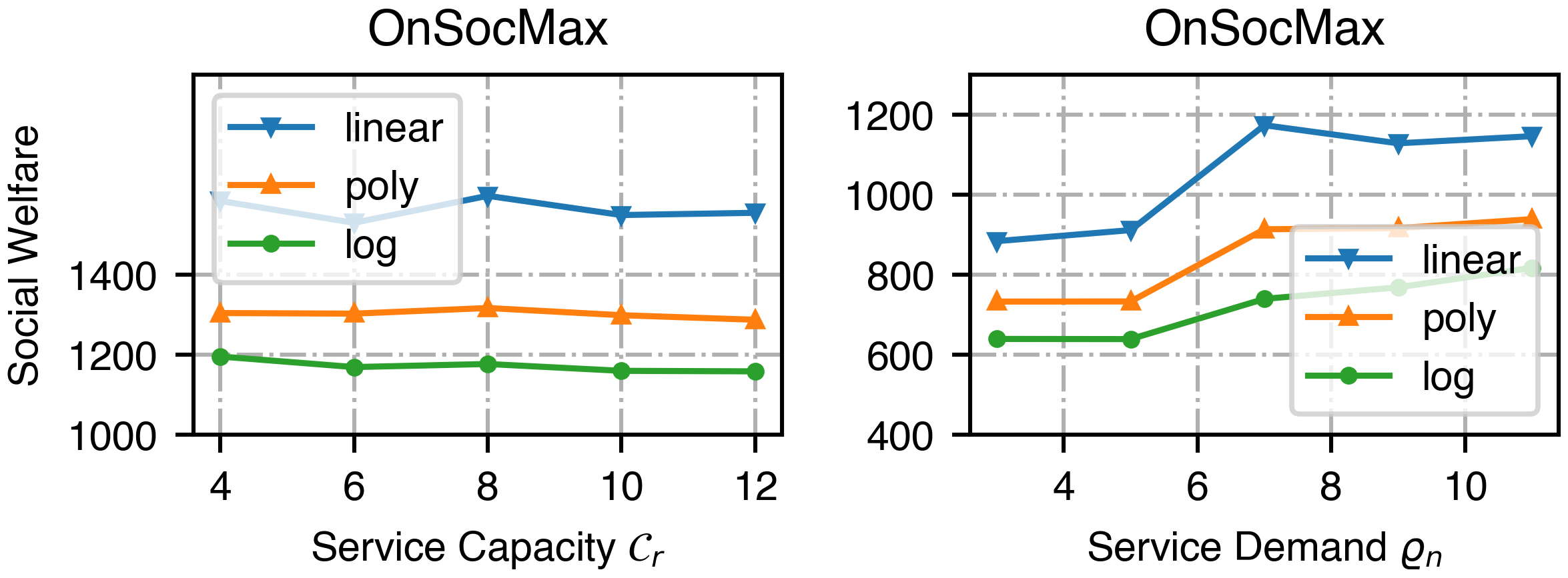}}
    \caption{Social welfare under different congestion levels. }
    \label{exp3}
\end{figure}

\textit{Robustness.} 
We verify the robustness of \textsc{OnSocMax} under different settings of maximum processable workload capacity of nodes and input workload size of jobs. These two 
variables actually tune the congestion level, i.e., the coverage rate of service demands, from different angles. We can conclude that \textsc{OnSocMax} 
is robust to the changes of the congestion level from Fig. \ref{exp3}.

\section{Related Work}\label{s5}
Workload dispatching and job scheduling are fully investigated under classic settings, where multiple identical nodes with exponentially distributed 
service rates process continuous arrived jobs. With Continuous Time Markov Chains (CTMC) and Lyapunov Stability theories, policies such as \textsc{JSQ} 
\cite{weber1978optimal}, \textsc{JIQ} \cite{10.1016/j.peva.2011.07.015}, \textsc{Pod} \cite{mukherjee2018universality}, and \textsc{JFIQ} \cite{weng2020optimal} are 
proposed and analyzed on the average response time and cross-server communication overhead. In a most recent work \cite{weng2020optimal}, 
Weng et al. proposed the \textsc{JFSQ} and \textsc{JFIQ} policies under the constraints of heterogenous service rates and service locality. They prove that, 
under a well-connected bipartite graph condition, these two policies achieve the minimum mean response time in both the 
\textit{many-server regime} and the \textit{sub Halfin-Whitt regime}. 

Another line of works study the energy-efficient and multi-resource sharing workload dispatching from a different theoretical basis 
\cite{energy-lb1,energy-lb2,energy-lb3,lb-queue,load-balancing1,ratio1,wan2018fog,8752939}. Generally, the objective is to improve the 
energy efficiency with on-demand resource allocation. Thereinto, online workload dispatching of deadline-aware (multi-server) jobs have 
been studied in \cite{energy-lb2,load-balancing1,ratio1}. In a similar work \cite{load-balancing1}, the authors design online algorithms 
for both fractional and non-fractional workload model under concave utility settings. The optimality of designed algorithms hold when 
all the jobs have the same deadline and share a single type of resource. Compared with it, our work is more general with the model of 
resource mesh and the technique of marginal cost estimation, which makes it more applicable to heterogeneous multi-server jobs with different 
deadlines. Online workload 
dispatching under the objective of minimizing the Nash Social Welfare is revisited in a recent paper \cite{simi}. This work provides 
tight bounds on the price of anarchy (PoA) of pure Nash equilibria and on the competitive ratio of the general greedy algorithm under 
very general latency functions. 

\section{Conclusion}\label{s6}
In this paper, we design an online workload dispatching policy for general multi-server jobs with the target of maximizing the social 
welfare. The multi-server jobs we considered have multiple task components and strict deadlines. To take both the spatio and temporal 
resource of the computing cluster into consideration, we establish a model of resource mesh. Each task component of jobs can only be 
dispatched to the nodes available to it based on the service locality constraints. With the marginal cost estimation technique, we design 
an online policy \textsc{OnSocMax} by solving several convex pseudo-social welfare maximization problems. The algorithm 
is proved to be $\alpha$-competitive for an $\alpha \geq 2$. Online workload dispatching of workflows with complex communication patterns 
will be studied in future.


\bibliographystyle{IEEEtran}
\bibliography{IEEEabrv,ref.bib}

\begin{thebibliography}{10}
\providecommand{\url}[1]{#1}
\csname url@samestyle\endcsname
\providecommand{\newblock}{\relax}
\providecommand{\bibinfo}[2]{#2}
\providecommand{\BIBentrySTDinterwordspacing}{\spaceskip=0pt\relax}
\providecommand{\BIBentryALTinterwordstretchfactor}{4}
\providecommand{\BIBentryALTinterwordspacing}{\spaceskip=\fontdimen2\font plus
\BIBentryALTinterwordstretchfactor\fontdimen3\font minus
  \fontdimen4\font\relax}
\providecommand{\BIBforeignlanguage}[2]{{%
\expandafter\ifx\csname l@#1\endcsname\relax
\typeout{** WARNING: IEEEtran.bst: No hyphenation pattern has been}%
\typeout{** loaded for the language `#1'. Using the pattern for}%
\typeout{** the default language instead.}%
\else
\language=\csname l@#1\endcsname
\fi
#2}}
\providecommand{\BIBdecl}{\relax}
\BIBdecl

\bibitem{nccl}
{NVIDIA Corporation}, ``The nvidia collective communication library,''
  \url{https://developer.nvidia.com/nccl}, 2022.

\bibitem{DL2}
Y.~Peng, Y.~Bao, Y.~Chen, C.~Wu, C.~Meng, and W.~Lin, ``Dl2: A deep
  learning-driven scheduler for deep learning clusters,'' \emph{IEEE
  Transactions on Parallel \& Distributed Systems}, vol.~32, no.~08, pp.
  1947--1960, aug 2021.

\bibitem{carrion2022kubernetes}
C.~Carri{\'o}n, ``Kubernetes scheduling: Taxonomy, ongoing issues and
  challenges,'' \emph{ACM Computing Surveys (CSUR)}, 2022.

\bibitem{choi2004fine}
K.~Choi, R.~Soma, and M.~Pedram, ``Fine-grained dynamic voltage and frequency
  scaling for precise energy and performance tradeoff based on the ratio of
  off-chip access to on-chip computation times,'' \emph{IEEE transactions on
  computer-aided design of integrated circuits and systems}, vol.~24, no.~1,
  pp. 18--28, 2004.

\bibitem{rountree2011practical}
B.~Rountree, D.~K. Lowenthal, M.~Schulz, and B.~R. De~Supinski, ``Practical
  performance prediction under dynamic voltage frequency scaling,'' in
  \emph{2011 International Green Computing Conference and Workshops}.\hskip 1em
  plus 0.5em minus 0.4em\relax IEEE, 2011, pp. 1--8.

\bibitem{8917749}
Z.~Han, H.~Tan, X.-Y. Li, S.~H.-C. Jiang, Y.~Li, and F.~C.~M. Lau, ``Ondisc:
  Online latency-sensitive job dispatching and scheduling in heterogeneous
  edge-clouds,'' \emph{IEEE/ACM Transactions on Networking}, vol.~27, no.~6,
  pp. 2472--2485, 2019.

\bibitem{gautam2015survey}
J.~V. Gautam, H.~B. Prajapati, V.~K. Dabhi, and S.~Chaudhary, ``A survey on job
  scheduling algorithms in big data processing,'' in \emph{2015 IEEE
  International Conference on Electrical, Computer and Communication
  Technologies (ICECCT)}.\hskip 1em plus 0.5em minus 0.4em\relax IEEE, 2015,
  pp. 1--11.

\bibitem{BSP}
Z.~Han, H.~Tan, S.~H.-C. Jiang, X.~Fu, W.~Cao, and F.~C. Lau, ``Scheduling
  placement-sensitive bsp jobs with inaccurate execution time estimation,'' in
  \emph{IEEE INFOCOM 2020 - IEEE Conference on Computer Communications}.\hskip
  1em plus 0.5em minus 0.4em\relax IEEE Press, 2020, p. 1053–1062.

\bibitem{8486422}
Y.~Bao, Y.~Peng, C.~Wu, and Z.~Li, ``Online job scheduling in distributed
  machine learning clusters,'' in \emph{IEEE INFOCOM 2018 - IEEE Conference on
  Computer Communications}, 2018, pp. 495--503.

\bibitem{attiya2020job}
I.~Attiya, M.~Abd~Elaziz, and S.~Xiong, ``Job scheduling in cloud computing
  using a modified harris hawks optimization and simulated annealing
  algorithm,'' \emph{Computational intelligence and neuroscience}, vol. 2020,
  2020.

\bibitem{zhang2020evolving}
F.~Zhang, Y.~Mei, S.~Nguyen, and M.~Zhang, ``Evolving scheduling heuristics via
  genetic programming with feature selection in dynamic flexible job-shop
  scheduling,'' \emph{ieee transactions on cybernetics}, 2020.

\bibitem{liang2020data}
S.~Liang, Z.~Yang, F.~Jin, and Y.~Chen, ``Data centers job scheduling with deep
  reinforcement learning,'' in \emph{Pacific-Asia Conference on Knowledge
  Discovery and Data Mining}.\hskip 1em plus 0.5em minus 0.4em\relax Springer,
  2020, pp. 906--917.

\bibitem{narayanan2020heterogeneity}
D.~Narayanan, K.~Santhanam, F.~Kazhamiaka, A.~Phanishayee, and M.~Zaharia,
  ``Heterogeneity-aware cluster scheduling policies for deep learning
  workloads,'' in \emph{14th $\{$USENIX$\}$ Symposium on Operating Systems
  Design and Implementation ($\{$OSDI$\}$ 20)}, 2020, pp. 481--498.

\bibitem{flow-time}
Z.~Hu, B.~Li, C.~Chen, and X.~Ke, ``Flowtime: Dynamic scheduling of
  deadline-aware workflows and ad-hoc jobs,'' in \emph{2018 IEEE 38th
  International Conference on Distributed Computing Systems (ICDCS)}, 2018, pp.
  929--938.

\bibitem{yarn}
Y.~Gao and K.~Zhang, ``Deadline-aware preemptive job scheduling in hadoop yarn
  clusters,'' in \emph{2022 IEEE 25th International Conference on Computer
  Supported Cooperative Work in Design (CSCWD)}, 2022, pp. 1269--1274.

\bibitem{map-reduce}
D.~Cheng, X.~Zhou, Y.~Xu, L.~Liu, and C.~Jiang, ``Deadline-aware mapreduce job
  scheduling with dynamic resource availability,'' \emph{IEEE Transactions on
  Parallel and Distributed Systems}, vol.~30, no.~4, pp. 814--826, 2019.

\bibitem{fairness}
T.~Lan, D.~Kao, M.~Chiang, and A.~Sabharwal, ``An axiomatic theory of fairness
  in network resource allocation,'' in \emph{2010 Proceedings IEEE INFOCOM},
  2010, pp. 1--9.

\bibitem{serverless2}
E.~Jonas, J.~Schleier-Smith, V.~Sreekanti, C.-C. Tsai, A.~Khandelwal, Q.~Pu,
  V.~Shankar, J.~Carreira, K.~Krauth, N.~Yadwadkar \emph{et~al.}, ``Cloud
  programming simplified: A berkeley view on serverless computing,''
  \emph{arXiv preprint arXiv:1902.03383}, 2019.

\bibitem{load-balancing1}
Z.~Zheng and N.~B. Shroff, ``Online multi-resource allocation for deadline
  sensitive jobs with partial values in the cloud,'' in \emph{IEEE INFOCOM 2016
  - The 35th Annual IEEE International Conference on Computer Communications},
  2016, pp. 1--9.

\bibitem{energy-lb2}
Z.~Liu, M.~Lin, A.~Wierman, S.~Low, and L.~L.~H. Andrew, ``Greening
  geographical load balancing,'' \emph{IEEE/ACM Transactions on Networking},
  vol.~23, no.~2, pp. 657--671, 2015.

\bibitem{zhang2018load}
J.~Zhang, F.~R. Yu, S.~Wang, T.~Huang, Z.~Liu, and Y.~Liu, ``Load balancing in
  data center networks: A survey,'' \emph{IEEE Communications Surveys \&
  Tutorials}, vol.~20, no.~3, pp. 2324--2352, 2018.

\bibitem{dpos-base}
X.~Tan, B.~Sun, A.~Leon-Garcia, Y.~Wu, and D.~H. Tsang, ``Mechanism design for
  online resource allocation: A unified approach,'' \emph{Proc. ACM Meas. Anal.
  Comput. Syst.}, vol.~4, no.~2, Jun. 2020.

\bibitem{dpos}
H.~Zhao, S.~Deng, Z.~Liu, Z.~Xiang, J.~Yin, S.~Dustdar, and A.~Zomaya, ``Dpos:
  Decentralized, privacy-preserving, and low-complexity online slicing for
  multi-tenant networks,'' \emph{IEEE Transactions on Mobile Computing}, pp.
  1--1, 2021.

\bibitem{network-utility}
T.~Lan, D.~Kao, M.~Chiang, and A.~Sabharwal, ``An axiomatic theory of fairness
  in network resource allocation,'' in \emph{2010 Proceedings IEEE INFOCOM},
  2010, pp. 1--9.

\bibitem{knapsack-multi}
J.~Puchinger, G.~R. Raidl, and U.~Pferschy, ``The multidimensional knapsack
  problem: Structure and algorithms,'' \emph{INFORMS Journal on Computing},
  vol.~22, no.~2, pp. 250--265, 2010.

\bibitem{assume1}
Z.~Zhang, Z.~Li, and C.~Wu, ``Optimal posted prices for online cloud resource
  allocation,'' \emph{Proc. ACM Meas. Anal. Comput. Syst.}, vol.~1, no.~1, Jun.
  2017.

\bibitem{ota}
Y.~Zhou, D.~Chakrabarty, and R.~Lukose, ``Budget constrained bidding in keyword
  auctions and online knapsack problems,'' in \emph{Internet and Network
  Economics}, C.~Papadimitriou and S.~Zhang, Eds.\hskip 1em plus 0.5em minus
  0.4em\relax Berlin, Heidelberg: Springer Berlin Heidelberg, 2008, pp.
  566--576.

\bibitem{assume3}
B.~Sun, A.~Zeynali, T.~Li, M.~Hajiesmaili, A.~Wierman, and D.~H. Tsang,
  ``Competitive algorithms for the online multiple knapsack problem with
  application to electric vehicle charging,'' \emph{Proc. ACM Meas. Anal.
  Comput. Syst.}, vol.~4, no.~3, Nov. 2020.

\bibitem{ev-charing1}
Z.~Zheng and N.~Shroff, ``Online welfare maximization for electric vehicle
  charging with electricity cost,'' in \emph{Proceedings of the 5th
  International Conference on Future Energy Systems}, 2014, p. 253–263.

\bibitem{ota-2}
X.~Tan, A.~Leon-Garcia, Y.~Wu, and D.~H.~K. Tsang, ``Online combinatorial
  auctions for resource allocation with supply costs and capacity limits,''
  \emph{IEEE Journal on Selected Areas in Communications}, vol.~38, no.~4, pp.
  655--668, 2020.

\bibitem{estimate-add}
L.~Yang, M.~H. Hajiesmaili, and W.~S. Wong, ``Online linear programming with
  uncertain constraints : (invited paper),'' in \emph{2019 53rd Annual
  Conference on Information Sciences and Systems (CISS)}, 2019, pp. 1--6.

\bibitem{competitive}
A.~Borodin and R.~El-Yaniv, \emph{Online computation and competitive
  analysis}.\hskip 1em plus 0.5em minus 0.4em\relax Cambridge University Press,
  2005.

\bibitem{gronwall}
D.~S. Mitrinovic, J.~Pecaric, and A.~M. Fink, \emph{Inequalities involving
  functions and their integrals and derivatives}.\hskip 1em plus 0.5em minus
  0.4em\relax Springer Science \& Business Media, 2012, vol.~53.

\bibitem{got}
R.~El-Yaniv, A.~Fiat, R.~M. Karp, and G.~Turpin, ``Optimal search and one-way
  trading online algorithms,'' \emph{Algorithmica}, vol.~30, no.~1, pp.
  101--139, 2001.

\bibitem{weber1978optimal}
R.~R. Weber, ``On the optimal assignment of customers to parallel servers,''
  \emph{Journal of Applied Probability}, pp. 406--413, 1978.

\bibitem{10.1016/j.peva.2011.07.015}
Y.~Lu, Q.~Xie, G.~Kliot, A.~Geller, J.~R. Larus, and A.~Greenberg,
  ``Join-idle-queue: A novel load balancing algorithm for dynamically scalable
  web services,'' \emph{Perform. Eval.}, vol.~68, no.~11, p. 1056–1071, Nov.
  2011.

\bibitem{mukherjee2018universality}
D.~Mukherjee, S.~C. Borst, J.~S. Van~Leeuwaarden, and P.~A. Whiting,
  ``Universality of power-of-d load balancing in many-server systems,''
  \emph{Stochastic Systems}, vol.~8, no.~4, pp. 265--292, 2018.

\bibitem{weng2020optimal}
W.~Weng, X.~Zhou, and R.~Srikant, ``Optimal load balancing in bipartite
  graphs,'' \emph{arXiv preprint arXiv:2008.08830}, 2020.

\bibitem{energy-lb1}
M.~A. Adnan, R.~Sugihara, and R.~K. Gupta, ``Energy efficient geographical load
  balancing via dynamic deferral of workload,'' in \emph{2012 IEEE Fifth
  International Conference on Cloud Computing}, 2012, pp. 188--195.

\bibitem{energy-lb3}
J.~Luo, L.~Rao, and X.~Liu, ``Temporal load balancing with service delay
  guarantees for data center energy cost optimization,'' \emph{IEEE
  Transactions on Parallel and Distributed Systems}, vol.~25, no.~3, pp.
  775--784, 2014.

\bibitem{lb-queue}
S.~Sthapit, J.~Thompson, N.~M. Robertson, and J.~R. Hopgood, ``Computational
  load balancing on the edge in absence of cloud and fog,'' \emph{IEEE
  Transactions on Mobile Computing}, vol.~18, no.~7, pp. 1499--1512, 2019.

\bibitem{ratio1}
B.~Lucier, I.~Menache, J.~S. Naor, and J.~Yaniv, ``Efficient online scheduling
  for deadline-sensitive jobs: Extended abstract,'' in \emph{Proceedings of the
  Twenty-Fifth Annual ACM Symposium on Parallelism in Algorithms and
  Architectures}, 2013, p. 305–314.

\bibitem{wan2018fog}
J.~Wan, B.~Chen, S.~Wang, M.~Xia, D.~Li, and C.~Liu, ``Fog computing for
  energy-aware load balancing and scheduling in smart factory,'' \emph{IEEE
  Transactions on Industrial Informatics}, vol.~14, no.~10, pp. 4548--4556,
  2018.

\bibitem{8752939}
G.~Aumala, E.~Boza, L.~Ortiz-Avilés, G.~Totoy, and C.~Abad, ``Beyond load
  balancing: Package-aware scheduling for serverless platforms,'' in \emph{2019
  19th IEEE/ACM International Symposium on Cluster, Cloud and Grid Computing
  (CCGRID)}, 2019, pp. 282--291.

\bibitem{simi}
V.~Bil{\`o}, G.~Monaco, L.~Moscardelli, and C.~Vinci, ``Nash social welfare in
  selfish and online load balancing,'' in \emph{Web and Internet Economics},
  X.~Chen, N.~Gravin, M.~Hoefer, and R.~Mehta, Eds.\hskip 1em plus 0.5em minus
  0.4em\relax Cham: Springer International Publishing, 2020, pp. 323--337.

\end{thebibliography}
\begin{IEEEbiography}
    [{\includegraphics[width=1in,height=1.25in,clip,keepaspectratio]{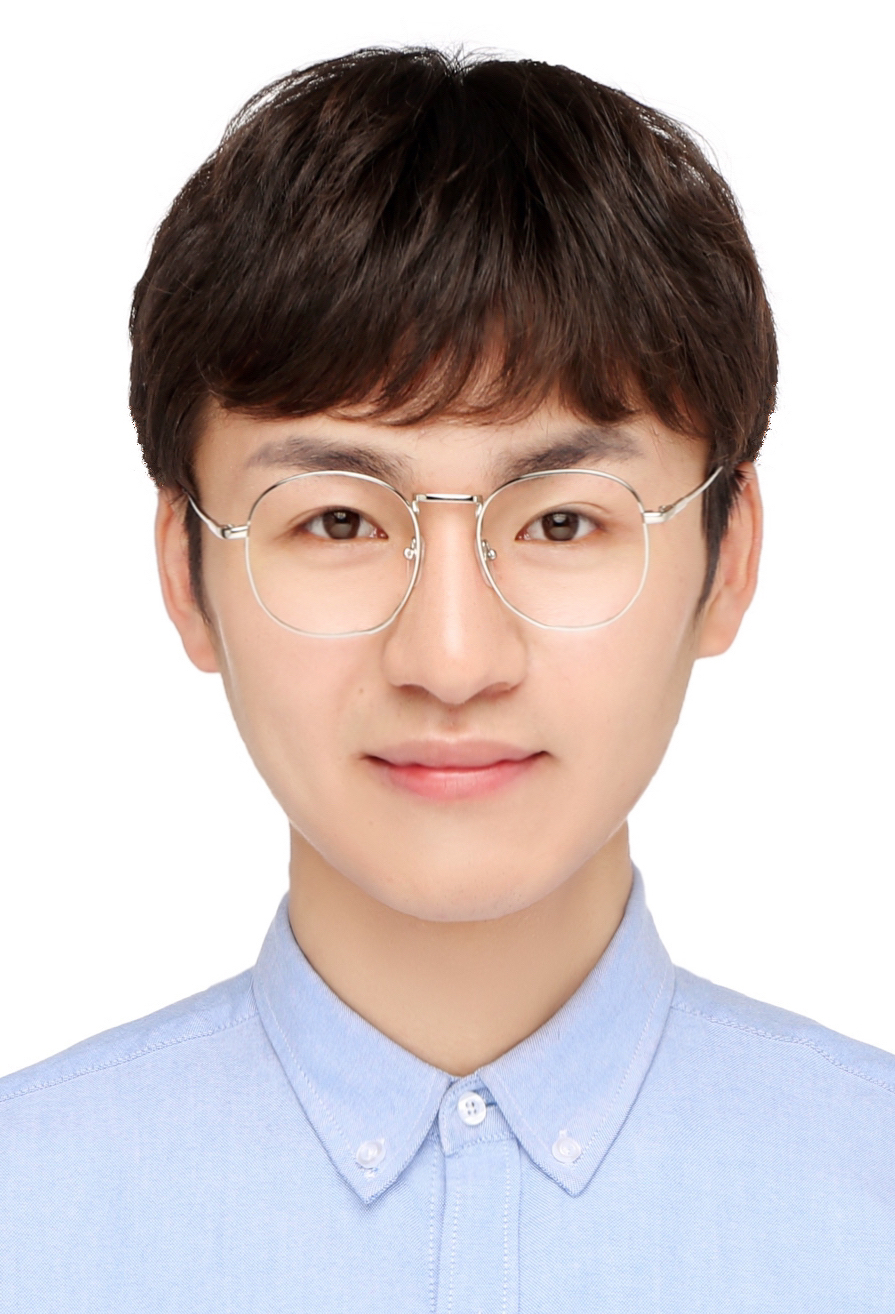}}]
    {Hailiang Zhao} received the B.S. degree in 
    2019 from the school of computer science and technology, Wuhan University of Technology, Wuhan, China. He is currently pursuing the 
    Ph.D. degree with the College of Computer Science and Technology, Zhejiang University, Hangzhou, China. His research interests include 
    cloud \& edge computing, distributed systems and optimization algorithms. He has published several papers in flagship conferences 
    and journals such as IEEE ICWS 2019, IEEE TPDS, IEEE TMC, etc. He was a recipient of the Best Student Paper Award of IEEE ICWS 2019. 
    He is a reviewer for IEEE TSC and Internet of Things Journal.
\end{IEEEbiography}

\begin{IEEEbiography}
    [{\includegraphics[width=1in,height=1.25in,clip,keepaspectratio]{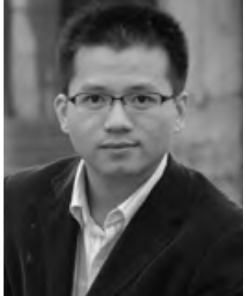}}]
    {Shuiguang Deng} 
    is currently a full professor at the College of Computer Science and Technology in Zhejiang University, China, 
    where he received a BS and PhD degree both in Computer Science in 2002 and 2007, respectively. He previously 
    worked at the Massachusetts Institute of Technology in 2014 and Stanford University in 2015 as a visiting scholar. 
    His research interests include Edge Computing, Service Computing, Cloud Computing, and Business Process Management. 
    He serves for the journal IEEE Trans. on Services Computing, Knowledge and Information Systems, Computing, and IET 
    Cyber-Physical Systems: Theory \& Applications as an Associate Editor. Up to now, he has published more than 100 
    papers in journals and refereed conferences. In 2018, he was granted the Rising Star Award by IEEE TCSVC. He is 
    a fellow of IET and a senior member of IEEE.
\end{IEEEbiography}

\begin{IEEEbiography}
    [{\includegraphics[width=1in,height=1.25in,clip,keepaspectratio]{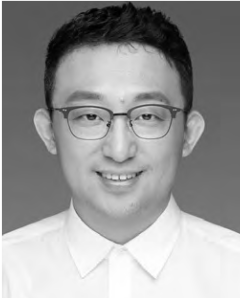}}]
    {Jianwei Yin} 
    received the Ph.D. degree in computer science from Zhejiang University (ZJU) in 2001. 
    He was a Visiting Scholar with the Georgia Institute of Technology. He is currently a Full Professor 
    with the College of Computer Science, ZJU. Up to now, he has published more than 100 papers in top 
    international journals and conferences. His current research interests include service computing 
    and business process management. He is an Associate Editor of the IEEE Transactions on Services 
    Computing.
\end{IEEEbiography}

\begin{IEEEbiography}
    [{\includegraphics[width=1in,height=1.25in,clip,keepaspectratio]{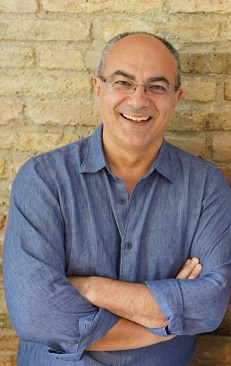}}]
    {Schahram Dustdar}
    is a Full Professor of Computer Science (Informatics) with a focus on Internet Technologies heading the Distributed 
    Systems Group at the TU Wien. He is founding co-Editor-in-Chief of ACM Transactions on Internet of Things (ACM TIoT) as well as Editor-in-Chief of Computing (Springer). He is an Associate Editor of IEEE Transactions on Services Computing, IEEE Transactions on Cloud Computing, ACM Computing Surveys, ACM Transactions on the Web, and ACM Transactions on Internet Technology, as well as on the editorial board of IEEE Internet Computing and IEEE Computer. Dustdar is recipient of multiple awards: TCI Distinguished Service Award (2021), IEEE TCSVC Outstanding Leadership Award (2018), IEEE TCSC Award for Excellence in Scalable Computing (2019), ACM Distinguished Scientist (2009), ACM Distinguished Speaker (2021), IBM Faculty Award (2012). He is an elected member of the Academia Europaea: The Academy of Europe, where he is chairman of the Informatics Section, as well as an IEEE Fellow (2016), an Asia-Pacific Artificial Intelligence Association (AAIA) President (2021) and Fellow (2021). He is an EAI Fellow (2021) and an I2CICC Fellow (2021). He is a Member of the 2022 IEEE Computer Society Fellow Evaluating Committee (2022).

\end{IEEEbiography}

\begin{IEEEbiography}
    [{\includegraphics[width=1in,height=1.25in,clip,keepaspectratio]{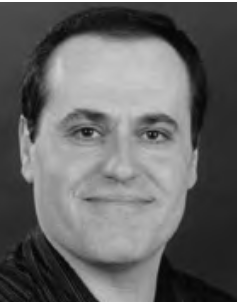}}]
    {Albert Y. Zomaya}
    is the Peter Nicol Russell Chair Professor of Computer Science and Director of the Centre for Distributed 
    and High-Performance Computing at the University of Sydney. To date, he has published > 600 scientific papers and articles and is (co-)author/editor 
    of > 30 books. A sought-after speaker, he has delivered > 250 keynote addresses, invited seminars, and media briefings. His research interests 
    span several areas in parallel and distributed computing and complex systems. He is currently the Editor in Chief of the ACM Computing Surveys 
    and served in the past as Editor in Chief of the IEEE Transactions on Computers (2010-2014) and the IEEE Transactions on Sustainable Computing (2016-2020).
    
    Professor Zomaya is a decorated scholar with numerous accolades including Fellowship of the IEEE, the American Association for the Advancement 
    of Science, and the Institution of Engineering and Technology (UK). Also, he is an Elected Fellow of the Royal Society of New South Wales and 
    an Elected Foreign Member of Academia Europaea. He is the recipient of the 1997 Edgeworth David Medal from the Royal Society of New South Wales 
    for outstanding contributions to Australian Science, the IEEE Technical Committee on Parallel Processing Outstanding Service Award (2011), 
    IEEE Technical Committee on Scalable Computing Medal for Excellence in Scalable Computing (2011), IEEE Computer Society Technical Achievement 
    Award (2014), ACM MSWIM Reginald A. Fessenden Award (2017), the New South Wales Premier’s Prize of Excellence in Engineering and Information 
    and Communications Technology (2019), and the Research Innovation Award, IEEE Technical Committee on Cloud Computing (2021). 
  \end{IEEEbiography}

\end{document}